\documentclass[aip,author-year]{revtex4-1}

\usepackage{amsmath,amssymb,amsfonts,amsthm}
\usepackage{graphicx}
\usepackage[english]{babel}

\usepackage{mathrsfs}
\newcommand{\vct}[1]{\mathbf{#1}}
\newcommand{\kb}[0]{k_{\textup{B}}}

\usepackage{float}

\usepackage[normalem]{ulem}
\usepackage{color}

\begin{document}

\title{Stress localization, stiffening and yielding in a model colloidal gel}
\author{Jader Colombo}
\affiliation{Department of Civil, Environmental and Geomatic Engineering,
  ETH Z\"{u}rich, CH-8093 Z\"{u}rich, Switzerland }
\author{Emanuela Del Gado}
\affiliation{Department of Civil, Environmental and Geomatic Engineering,
  ETH Z\"{u}rich, CH-8093 Z\"{u}rich, Switzerland }
\affiliation{Institute for Soft Matter Synthesis and Metrology \& Department of Physics, Georgetown University, 20057 Washington DC, USA}
\begin{abstract}
We use numerical simulations and an athermal quasi-static shear protocol to investigate the yielding of a model colloidal gel. Under increasing deformation, the elastic regime is followed by a significant stiffening before yielding takes place. A space-resolved analysis of deformations and stresses unravel how the complex load curve observed is the result of stress localization and that the yielding can take place by breaking a very small fraction of the network connections. {The stiffening corresponds to the stretching of the network chains, unbent and aligned along the direction of maximum extension. It is characterized by a strong localization of tensile stresses, that triggers the breaking of a few network nodes at around 30$\%$ of strain. Increasing deformation favors further breaking but also shear-induced bonding, eventually leading to a large-scale reorganization of the gel structure at the yielding. At low enough shear rates, density and velocity profiles display significant spatial inhomogeneity during yielding in agreement with experimental observations.}
\end{abstract}
\pacs{}
\maketitle

\section{Introduction}
The non-linear mechanical response of gels is of great relevance to technological applications requiring unusual material properties, such as adaptability or self-healing \citep{sottos_nature,leibler_nature,pochan_softmatter2010}, and has a crucial role in the functions of bio-materials and biological systems \citep{lieleg_nmat2011,storm-nature2005}. It is the result of the non-trivial stress transmission through the gel network structure, that can allow for stress or strain localization \citep{schall-science2007}, as well as complex spatial correlations, and it is far from being understood and controlled. 
Colloidal gels, that can form in suspensions of colloidal particles in the presence of attractive effective interactions, are particularly appealing as materials whose functions can be in principle designed at the level of the nanoscale (particle) components \citep{dimichele2013,sacanna2013}. In colloidal suspensions, gels can form even in extremely dilute systems via aggregation of the particles into a rich variety of network structures, that can be suitably tuned by changing the solid volume fraction, the physico-chemical environment or the processing conditions. Hence these {\it handles} could be used to design a specific complex mechanical response in addition to the selected nano-particle properties \citep{gibaud-prl2013,conrad-jor2010}. 

{Colloidal gels are typically very soft, but the variety of microstructures may lead to an equal variety in the mechanics, with elastic moduli spanning from $ \simeq 0.1$ up to $\simeq 100$ Pa \citep{gisler1999,exp2,zaccone_prl2009,Helgeson-softmat2014}. The microstructural complexity may also enable adjustments of the mechanical response to the external deformation. Soft gels can be in principle made to yield relatively easily, but in certain cases a significant strain hardening has been observed before yielding finally occurs \citep{gisler1999,Pouzot-jcis2006}. The extended non-linear behavior, reminiscent of the strain stiffening of biopolymer networks \citep{storm-nature2005,gardel-science2004,blair-collagen2010}, has been suggested to arise from the nonlinear response to elongation of the part of the colloidal gel structure that is responsible for stress-bearing.  

The yielding of colloidal gels, due to breaking and reorganization of the network structure, can be accompanied by strong inhomogeneity of stresses and strains throughout the material \citep{picard-pre2005,Mohraz-jor2005,vermant_jor2009,ovarlez-rheoacta2009,fall-prl2010,rajaram_sm2010,fielding-arxiv2013}. In particular, the stress-bearing backbone of the gels can be relatively poorly connected or floppy, hence the imposed deformation can induce significant and non-trivial structural rearrangements prior to and during yielding. The stress necessary to yield and the amount of plastic deformation accumulated in the material upon yielding are fundamental quantities to control. They may be crucially affected by the gel restructuring, as well as by the strain rate, in a way that is far from being rationalized \citep{Denn-rheoacta2011}. In this respect, most of theoretical and numerical works have focused on dense systems \citep{chaudhuri-pre2012,barrat-lemaitre,martens-prl2011} and diluted gels have been hardly touched upon. In spite of a certain insight gained in various experiments into the phenomena associated to the non-linear mechanical response of low-density colloidal gels, further understanding has been severely hindered because information on local deformations or structural rearrangements is very difficult to obtain experimentally.  Most of existing numerical studies on gel mechanics deal only with the linear response regime \citep{lodge-jor99,santos-sm2013}. Numerical methods to investigate non-linear phenomena are just under development \citep{swan-jor2013} and even recent studies are limited to testing specific deformation mechanisms on artificially built structures \citep{seto-jor2013,lindstrom-sm2012}, because microscopic models able to investigate the deformation behaviour in the presence of the same forces that led to the gel structure through aggregation are fundamentally lacking.}

Here we investigate the non-linear response of a colloidal gel at low volume fraction by numerical simulations of a microscopic model, that allows us to analyze the response to deformation arising from the network  structure and from the same effective interactions that have driven its self-assembly. We have designed an athermal quasi-static deformation protocol, in the same spirit of recent numerical studies of deformation in amorphous solids \citep{tanguy2002continuum,maloney2006amorphous,falk-langer,procaccia,fiocco-pre2013}, that allows us to compute the load curve of the material at relatively low strain rates. By combining this approach with a space-resolved analysis of microscopic processes, deformations and stresses, we gain significant new insight into the non-linear response and the yielding of the gel. 

\section{Model and Numerical Simulations}\label{sec:numsim}

{Gelation in diluted colloidal suspensions may be the result of an arrested phase separation, spinodal decomposition, or of a diffusion limited aggregation \citep{Carpineti-prl1992,plu-nature,Gibaud-jpcm2009,Eberle-prl2011}. To capture all the complexity of these phenomena is challenging and a number of particle-based models for gel formation have been devised in the last few years including short-range isotropic interactions~\citep{charbonneu2007_pre,tanaka2007_epl,foffi2005_jcp}, valence-limited and patchy-particle models \citep{kern-frenkel,sciortino,bianchi} and anisotropic effective interactions \citep{edgkob_epl05,saw2009structural,delgado2010microscopic}. 

We follow here this latter approach, using an effective interaction that includes, in the form of a three-body term, the basic ingredients for a minimal model of particle gels. In fact dilute colloidal gels are characterized by open and thin network structures, where particle coordination can be very low, hence the network connections need to be fairly rigid to support at least their own weight. Experiments have proven that bonds between the colloidal particles can indeed support significant torques~\citep{solomon,royall, pantina2005elasticity,dinsmore_prl2006}. In our model at low volume fractions the particles self-assemble into a thin open structure, mechanically stable and locally rigid thanks to the anisotropic interactions \citep{colombo_prl2013,colombo-sm2014}. The relatively simple gel structure certainly does not capture all the structural complexity of these materials but it allows us to rationalize the connection between structure, local processes and mechanics. In this respect we consider that the particles in our model gel can be thought of as single colloidal particles in the case of very diluted {\it stringy} gels or else as {\it chunks} or aggregates that are assembled into the gel structure in the thicker gels which typically result from phase separation.} 

Our system consists of $N$ identical particles with position vectors $\{\vct{r}_i\}\,,i=1\ldots N$, interacting via the potential energy
\begin{equation}\label{equ:poten}
U(\vct{r}_1,\ldots,\vct{r}_N) = \epsilon \left[\sum_{i > j} u_2\left(\frac{\vct{r}_{ij}}{\sigma}\right) +
 \sum_i\sum_{\substack{j>k}}^{j,k\ne i}u_3\left(\frac{\vct{r}_{ij}}{\sigma},\frac{\vct{r}_{ik}}{\sigma}\right)\right]\,,
\end{equation}
where $\vct{r}_{ij}=\vct{r}_j-\vct{r}_i$, $\epsilon$ sets the energy scale and $\sigma$ represents
the particle diameter. Typical values for a colloidal system are $\sigma = 10-100\,\rm{nm}$ and
$\epsilon = 1-100\,\kb T$, $\kb$ being the Boltzmann constant and $T$ the room
temperature~\citep{trappe_nature2001,luca_faraday2003,laurati_jor2011}. The two-body term $u_2$
consists of a repulsive core complemented by a narrow attractive well:
\begin{equation}\label{equ:u2}
  u_2(\vct{r})=A\left(a\,r^{-18}-r^{-16}\right)\,.
\end{equation}
The three-body term $u_3$ confers angular rigidity to the inter-particle bonds: 
\begin{equation}
  u_3(\vct{r},\vct{r}') = B\,\Lambda(r) \Lambda(r')\,
  \exp\left[-\left(\frac{\vct{r}\cdot\vct{r}'}{rr'}-\cos\bar{\theta}\right)^2/w^2\right]\,.
\end{equation}
The range of the three-body interaction is equal to two particle diameters, as ensured by the radial
modulation
\begin{equation}
\Lambda(r)=
\begin{cases}
r^{-10} \left[1-(r/2)^{10}\right]^2 & r<2 \\
0 & r\ge 2
\end{cases}
\end{equation} 
The potential energy~\eqref{equ:poten} depends parametrically on the dimensionless quantities $A$, $a$, $B$, $\bar{\theta}$, $w$. We have chosen these parameters such that for $\kb T \sim 10^{-1}\epsilon$ the particles start to self-assemble into a persistent particle network. The data here discussed refer to $A=6.27$, $a=0.85$, $B=67.27$, $\bar{\theta}=65^\circ$, $w=0.30$, one convenient choice to realize this condition.

{We have extensively investigated the self-assembly of the gel at rest and performed a spatially resolved analysis of its cooperative microscopic dynamics in previous works \citep{colombo_prl2013,colombo-sm2014}. At low volume fractions ($0.05 \leq \Phi \leq 0.2$) the network consists of chains of particles connected by crosslinks (or {\it nodes}, i.e.\ the branching points) as, due to local rigidity, the particles form mainly $2$ or $3$ bonds each. Density fluctuations do not show any sign of a phase separation close to gelation. Nevertheless, the gel is characterized by the coexistence of poorly connected regions, i.e.\ regions with a lower local density of crosslinks, where major structural rearrangements tend to take place, and densely connected domains (i.e.\ regions where the local density of crosslinks is higher) where internal stresses tend to concentrate. We have shown that local bond breaking processes, driven by thermal fluctuations and relaxation of internal stresses, have non-local consequences and induce structural rearrangements relatively far away along the gel network structure.} The structural heterogeneity, the long range spatial correlations underlying the cooperative dynamics and the heterogeneous distribution of internal stresses in our model capture some fundamental physical characteristics of colloidal gel networks \citep{maccarrone}, that probably have a major role in their complex mechanical response. Here we investigate the behavior of this model gel network under deformation, from the linear elastic behavior to the gel yielding.  

All simulations were performed using the LAMMPS molecular dynamics source code~\citep{plimpton1995fast}, which we have suitably extended to include the interaction~\eqref{equ:poten}. The gel consists of $N=5\cdot10^5$ particles in a cubic simulation box with linear size $L\approx135\sigma$ such that the dimensionless number density $N\sigma^3/L^3$ equals $0.20$. This corresponds to an approximate volume fraction of $10\%$. We use periodic boundary conditions in the Lees-Edwards formulation~\citep{leesedwards1972}, which is compatible with shear deformations.

{The starting point of our study is a gel configuration equilibrated at $\kb T=5\cdot10^{-2}\epsilon$, whose structure and relaxation dynamics have been already thoroughly characterized. At this temperature thermal fluctuations alone may cause breaking of single inter-particle bonds, although they are not sufficient to destroy the gel structure within a reasonably large simulation time window. Hence, when the material is subjected to shear deformation it displays an elastic response that is only very weakly dependent on the shear rate, within a relatively wide range of shear rates \citep{colombo_prl2013}. 

Here we investigate the behaviour of the gel at large deformation strains, beyond the linear response regime. In particular we want to unravel the interplay between the internal stresses and the imposed deformation. In order to focus on the microscopic processes which are induced by the deformation and to distinguish them from thermal processes, we consider that the depth $\epsilon$ of the attractive potential well acting on the particles bonded in the network is much larger than the thermal energy $k_{B}T$. To capture these conditions  we perform athermal simulations (i.e., at zero temperature) using the following damped molecular dynamics: }
\begin{equation}\label{equ:equmotion}
  m\frac{d^2{\vct{r}}_i}{dt^2} = -\xi\frac{d{\vct{r}}_i}{dt} - \nabla_{\vct{r}_i}U
\end{equation}
where $m$ is the particle mass and $\xi$ the coefficient of friction. As a first step we quench the initial gel configuration (i.e., the one equilibrated at finite temperature) down to zero temperature by running a simulation with the damped dynamics~\eqref{equ:equmotion} until the kinetic energy drops to a negligible fraction (less than $10^{-10}$) of its initial value. The resulting configuration is a local minimum of the potential energy, or \emph{inherent structure}~\citep{stillinger-weber}.  
Starting from the inherent structure we perform a series of incremental strain steps in simple shear geometry. Each step increases the cumulative shear 
strain by a quantity $\delta\gamma$ and comprises two phases. During the first phase we apply an instantaneous affine deformation $\Gamma_{\delta\gamma}$,
corresponding to simple shear in the $xy$ plane, to all particles:
\begin{equation}\label{equ:ss1}
  \vct{r}_i' = \Gamma_{\delta\gamma}\vct{r}_i = 
  \begin{pmatrix}
    1 & \delta\gamma & 0 \\
    0 & 1 & 0 \\
    0 & 0 & 1
  \end{pmatrix}
  \vct{r}_i\,.
\end{equation}
The Lees-Edwards boundary conditions are updated as well to comply with the increase in the cumulative strain. During the second phase of the strain step we relax the affinely deformed configuration by letting the system free to evolve in time while keeping the global 
strain constant:
\begin{equation}\label{equ:ss2}
  \vct{r}_i'' = \mathscr{T}_{\delta t}\vct{r}_i'\,;
\end{equation}
$\mathscr{T}_{\delta t}$ is the time evolution operator for the damped dynamics~\eqref{equ:equmotion} and a specified time interval $\delta t$. 

Upon repeating the two phases for $n$ steps, the cumulative strain equals $\gamma=n\,\delta\gamma$ and the gel configuration is
\begin{equation}
  \vct{r}_{i,n} = (\mathscr{T}_{\delta t}\Gamma_{\delta\gamma})^n\,\vct{r}_{i,0}\,,
\end{equation}
where $\{\vct{r}_{i,0}\}$ denotes the configuration of the starting inherent structure. 

The procedure just outlined corresponds to the \emph{athermal quasistatic} (AQS) approach extensively used to investigate the deformation behavior of 
amorphous solids~\citep{tanguy2002continuum,maloney2006amorphous}, with the following difference: while in the 
quasistatic case the relaxation of the affinely deformed configuration is performed directly in the potential energy landscape via an energy minimization,
here we follow a more natural dynamics of the system (with viscous energy dissipation) for a prescribed time interval $\delta t$. As a consequence, we work at 
a finite shear rate $\dot{\gamma} = \delta\gamma/\delta t$, while the quasistatic case corresponds to the limit $\delta t\to\infty$, i.e.\ $\dot{\gamma}\to 0$. In our model 
system direct energy minimization schemes such as the steepest descent or conjugate gradient methods are not very effective, presumably because the 
potential energy landscape is characterized by extended relatively flat regions corresponding to floppy deformation modes of the gel \citep{alexander, wyart-prl2008, rovigatti2011}.

Disregarding effects due to the particle inertia, the microscopic dynamics~\eqref{equ:equmotion} introduce a natural time scale $\tau_0=\xi\sigma^2/\epsilon$, corresponding to the time it takes a particle subjected to a typical force $\epsilon/\sigma$ to move a distance equal to its size. {This is the relevant unit time scale here, since we refer to a system where $\epsilon \gg k_{B}T$ and the characteristic time for particle diffusion is much longer than the simulation time window. Hence we express times and rates referring to this natural unit of time. Along this line, $\tau_0$ can be used to define a modified P\'eclet number $\tilde{Pe}=\tau_0\dot{\gamma}$, that quantifies the possibility for the gel particles to rearrange under the effect of internal forces while being sheared at a shear rate $\dot{\gamma}$.

The time step for the integration of the equation of motion equals $10^{-2}\tau_0$. 
Fixing the elementary strain increment $\delta\gamma=10^{-2}$, we shear the gel for $n=100$ strain steps up to a final strain $\gamma=1.0$. We vary the relaxation interval $\delta t$ in order to 
obtain three different values of the effective shear rate: $\dot{\gamma}_1=2\cdot10^{-5}\,\tau_0^{-1}$, $\dot{\gamma}_2=10^{-4}\,\tau_0^{-1}$, and $\dot{\gamma}_3=10^{-3}\,\tau_0^{-1}$. In a typical aqueous solution of colloidal particles with a diameter $\sigma\approx 100$ nm and an interaction energy $\epsilon\approx 10\kb T$~\citep{petekidis_softmatter2011} the characteristic time is $\tau_0\approx 10^{-4}$ s; in such a system the three shear rates we have investigated 
would correspond to 0.2 $\rm{s}^{-1}$, 1.0 $\rm{s}^{-1}$ and 10 $\rm{s}^{-1}$, respectively  (in all cases the modified P\'eclet number $\tilde{Pe}\ll 1$). }

At the end of each strain step we compute the global stress tensor $\sigma_{\alpha\beta}$
($\alpha,\beta$ stand for the cartesian components $\{x,y,z\}$) using the virial definition in the
formulation introduced by~\citet{thompson2009general}, which is valid for arbitrary many-body
potentials in the presence of periodic boundary conditions. Since the velocities of the particles
are small ($v_i\lesssim 10^{-5} \sigma/\tau_0$ for all shear rates) we ignore in the calculation of
the stress tensor the kinetic term $mv_i^{\alpha}v_i^\beta$, as well as any effect of the viscous drag appearing in Eq.~\eqref{equ:equmotion}. Moreover, in order to characterize the spatial stress
distribution inside the gel we divide the simulation box in cubical sub-cells, and associate to each
cell $k$ a local stress tensor $\sigma^k_{\alpha\beta}$. The local stresses are defined by grouping
the contributions to the global virial in space: each cell receives the contributions of the
interactions involving the particles located inside the cell. This is a convenient way of estimating
the spatial variation of the state of stress in the material~\citep{thompson2009general}. We divide
the box in 40x40x40 cells, so that each cell contains on average about eight particles.

During the shearing we also monitor the degree of non-affinity of the microscopic particle displacements, i.e.\ the extent to which the local deformation departs from the one imposed at a global level. If the deformation were completely affine, at the end of $n$-th strain step the position vector of particle $i$ would correspond to $\Gamma_{\delta\gamma}^n \vct{r}_{i,0}$; we therefore introduce the \emph{average (per particle) non-affine displacement} $\Delta_n$, as a measure of the non affinity accumulated after $n$ steps:
\begin{equation}
  \Delta_n = \frac{1}{N}\sum_{i=1}^N \|\vct{r}_{i,n} - \Gamma_{\delta\gamma}^n \vct{r}_{i,0}\|\,.
\end{equation}

With the approach just described we obtain the load curve of the material at different shear rates and analyze the underlying microscopic processes, in terms of local stresses and strains and of structural modifications of the gel.

\section{Load curve: stiffening and yielding}
\label{sec:loadcurve}
\begin{figure}
  \includegraphics[width=.9\columnwidth, clip]{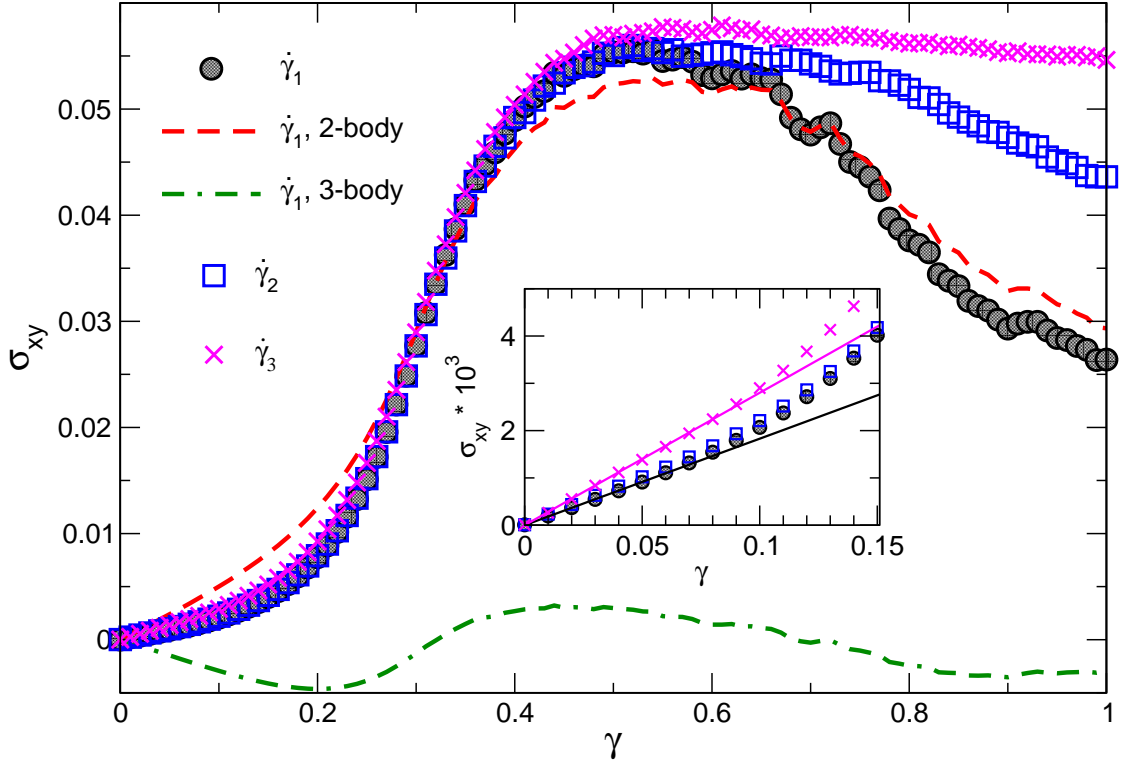}
  \caption{\label{fig:loadcurve} (Main plot) Load curves (average shear stress
      $\sigma_{xy}$ vs. shear strain $\gamma$) in a gel network with $5\cdot10^5$ particles, for
      three different shear rates: $\dot{\gamma}_1$ (low), $\dot{\gamma}_2$ (intermediate) and
      $\dot{\gamma}_3$ (high). The broken lines represent a partitioning of the total shear stress
      into the contribution of the 2-body potential (dashed line) and the 3-body potential
      (dashed-dotted line), respectively, for the case of slow shearing. (Inset) Magnified view of
      the load curves in the range $0\le\gamma\le0.15$. The straight lines represent linear fits to
      the small-deformation data for the low and high shear rate: the corresponding shear moduli are
      $1.8\cdot10^{-2}$ and $2.8\cdot10^{-2} \epsilon/\sigma^{3}$, respectively.}
\end{figure}

In Fig.~\ref{fig:loadcurve} we plot the load curve of the gel, i.e.\ the average stress as a
function of the imposed cumulative strain obtained in a system of $5\cdot10^5$ particles with the
procedure described above, for the three shear rates $\dot{\gamma}_1$, $\dot{\gamma}_2$ and
$\dot{\gamma}_3$ (indicated with the circles, squares and crosses, respectively). The data show that
a first linear elastic behavior is followed, for deformations larger than $10\%$, by a stiffening of
the material. The stress increase with $\gamma$ evolves toward a maximum around $\gamma \simeq 0.5$, which we define as a yielding point: after
  yielding the stresses progressively decrease with the imposed deformation.
This type of load curve is qualitatively very similar to the ones typically measured in colloidal
gels \citep{laurati_jor2011,Pouzot-jcis2006,gisler1999}. It is also reminiscent, to a certain extent, of the behaviour characterising different biopolymer networks, from collagen to actin \citep{storm-nature2005}.

 {The elastic regime extends over a maximum of $\simeq 10\%$ of strains (see inset of
  Fig.~\ref{fig:loadcurve}), in agreement with experimental observations. We can
  extract a small, but finite shear modulus ($G \simeq 2\cdot 10^{-2}\epsilon/\sigma^3$), which
  increases mildly for the highest shear rate}. For a typical colloidal gel where the particle diameter $\sigma\approx 100$ nm and the interaction energy $\epsilon\approx 10\kb T_{\rm r} $, this
corresponds to a shear modulus $G \simeq 1$ Pa, which is consistent with what measured in
experiments on a few dilute colloidal gel networks \citep{petekidis_softmatter2011,laurati_jor2011, gisler1999}.  We have distinguished in
Fig.~\ref{fig:loadcurve} the contributions to stresses due to the two- and three-body interactions
(respectively dashed and dashed-dotted lines) in the case of the low shear rate
$\dot{\gamma}_1$. The data show that the main contribution opposing the shear forces is due to the
two-body, cohesive forces, as expected. The contribution of the bending forces to the stresses, instead, is negative in sign for $\gamma \leq 0.3$ and becomes positive for $0.3 < \gamma <
0.7$. Hence for small and moderate deformation, at the beginning of the stiffening regime, bending forces tend to push in the direction of the shear: this could be due to the unbending of initially
compressed chains of the gels along the shear direction. Around $\gamma \simeq 0.3$, instead, the
bending forces start to resist to shear and their contribution increases with deformation up to the
end of the stiffening regime, suggesting that for higher deformation, until the material yields,
most of the chains are stretched out. Therefore, at this point, the bending forces should mainly
oppose the shear deformation acting at the level of the network crosslinks. {In addition, the plot also shows that the 2-body contribution to the stress becomes clearly predominant with the onset of the non-linear regime, suggesting that the stiffening we detect is directly related to the stretching of the extensible interparticle bonds.} This possible scenario points to further similarities between the mechanical response of our colloidal gel and the one of biopolymer networks, where the stiffening regime is ascribed to the shear deformation pulling on the stretched out biopolymers and has been also related to the presence of negative normal stresses \citep{jamney-natmat2007}.  Our gel is indeed under tension along the direction that corresponds to
the velocity gradient under shear ($y$), hence it would contract if volume changes were allowed in
the simulations. In Fig.~\ref{fig:loadcurve} we can also compare the data obtained at different
shear rates: within the range investigated here, the elastic and stiffening regimes do not
significantly vary with varying the shear rate, whereas the yielding regime is significantly
modified. In particular, the higher the shear rate, the higher the level of the shear stresses kept
in the material upon yielding. This finding is consistent with the yielding arising from the
development of plastic processes in the materials, whose amount and/or extent is affected by the
shear rate of the deformation applied.

{In most cases the experimental investigations we are referring to are based on oscillatory rheology. We have therefore tested the picture just obtained through the new numerical protocol proposed here by performing a set of simulations under oscillatory conditions. The details of the oscillatory rheology study are reported in Appendix \ref{appendix1} and its results are in good agreement with the ones discussed so far. In addition to the linear regime, we have also investigated the non-linear one following \citep{Hyun-pps2011,vanderVaart-jor2013,Ewoldt-jor2008} and computed the third-order viscoelastic coefficients as well as a set of Lissajous-Bowditch plots. This quantitative analysis confirms that the increase of the stresses in the non-linear regime is dominated by the elastic contributions and that the material is strain stiffening.} 

\begin{figure}
  \includegraphics[width=.9\columnwidth, clip]{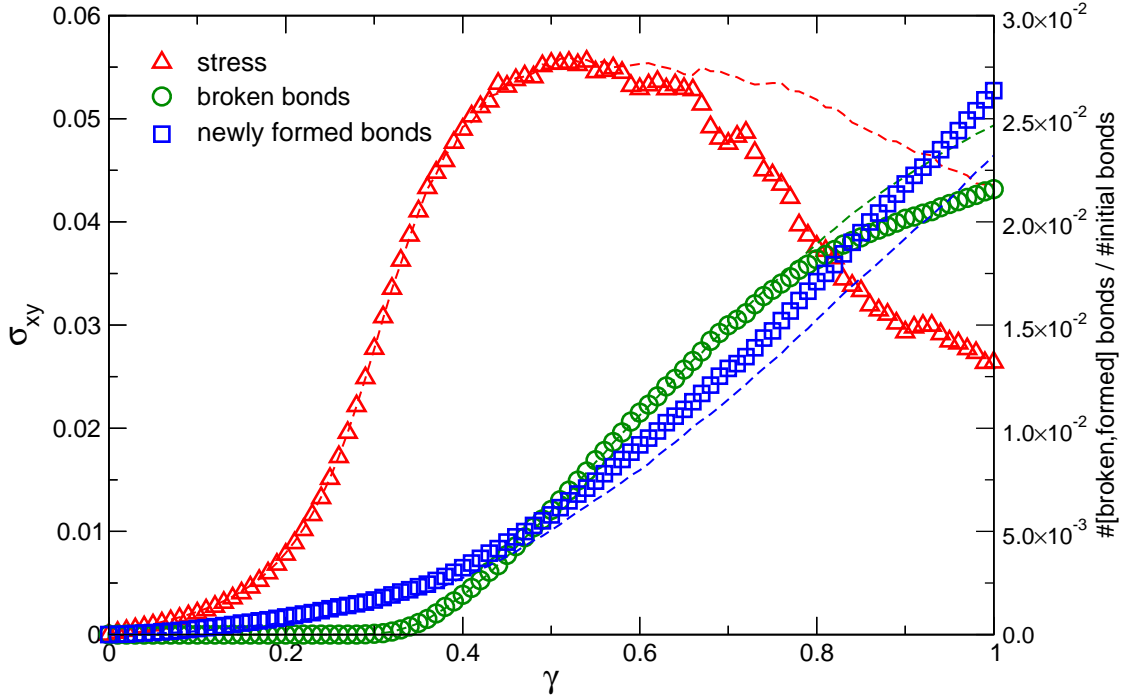}
  \caption{\label{fig:breakbond} Breaking of existing bonds and formation of new bonds in the gel (given as a fraction of the initial number of bonds) due to shear deformation, plotted together with the load curve. The symbols refer to the lowest shear rate $\dot{\gamma}_1$, whereas the dashed lines refer to the intermediate
   rate $\dot{\gamma}_2$.}
\end{figure}

{With the aim of investigating the microscopic interplay between deformation and internal stresses underlying the load curve, we have monitored in our simulations the bond breaking and formation.} In Fig.~\ref{fig:breakbond} we plot the total fraction of bonds broken and formed together with the load curve for the lowest ($\dot{\gamma}_1$) and intermediate ($\dot{\gamma}_2$) shear rates. 
One can immediately recognize that up to $\gamma \leq 30\%$, i.e.\ well into the non-linear stiffening regime, the deformation does induce some bond formation (although very limited) but does not break any bond. This is consistent with the picture sketched above, where the onset of the stiffening must be related to the shear deformation pulling on the stretched network chains. Beyond $\sim 30\%$, the deformation starts to break bonds: because $\gamma \geq 30\%$ also corresponds to the change in sign of the contribution to the shear stress of the bending forces (see Fig.~\ref{fig:loadcurve}), this finding suggests that the bonds that are first broken have a specific role in stress transmission within the gel structure. Finally, around $\gamma\sim 50\%$, once the fraction of broken bonds becomes of the order of,  or overcomes, the fraction of newly formed ones, the yielding process sets in. When comparing the data obtained with the different shear rates, one can see that yielding reached at lower $\dot{\gamma}$ is characterised by lower shear stresses and corresponds to comparatively less broken bonds and more newly induced ones. That higher shear rates deformation may break more bonds is to be expected, but that lower levels of shear stress, obtained upon yielding at lower shear rate, correspond to comparatively more bonds formed is somewhat counter-intuitive: one would rather think, in fact, that forming more bonds would allow to recover part of the network strength and to sustain higher stresses. The picture emerging from our data is therefore that bonds formed upon yielding play quite a different role for stress bearing with respect to the gel connections at rest, suggesting that a significant reorganization of the gel structure is taking place.

In order to unravel the microscopic origin of the mechanical behavior of our model colloidal gel and test the scenarios hypothesized above, in the following we exploit further the information contained in the simulation data and perform a space-resolved analysis of the stresses and the deformations in the material.

\section{Space-resolved analysis of stresses and deformations}
\label{sec:spaceresolved}
\begin{figure}
  \includegraphics[width=.9\columnwidth, clip]{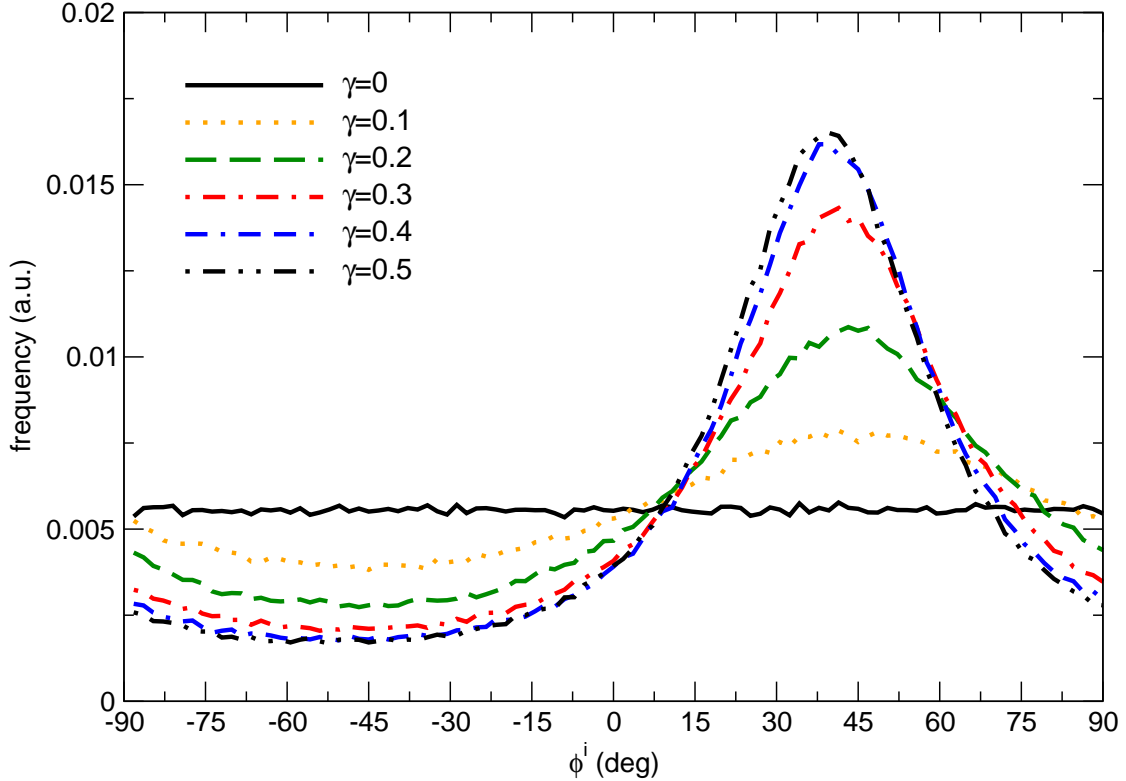}
  \caption{\label{fig:angle1} Histogram of the angles $\phi^i$ quantifying the progressive alignment of the shear stresses, upon increasing $\gamma$ up to $50\%$.}
\end{figure}

\begin{figure}
  \includegraphics[width=.9\columnwidth, clip]{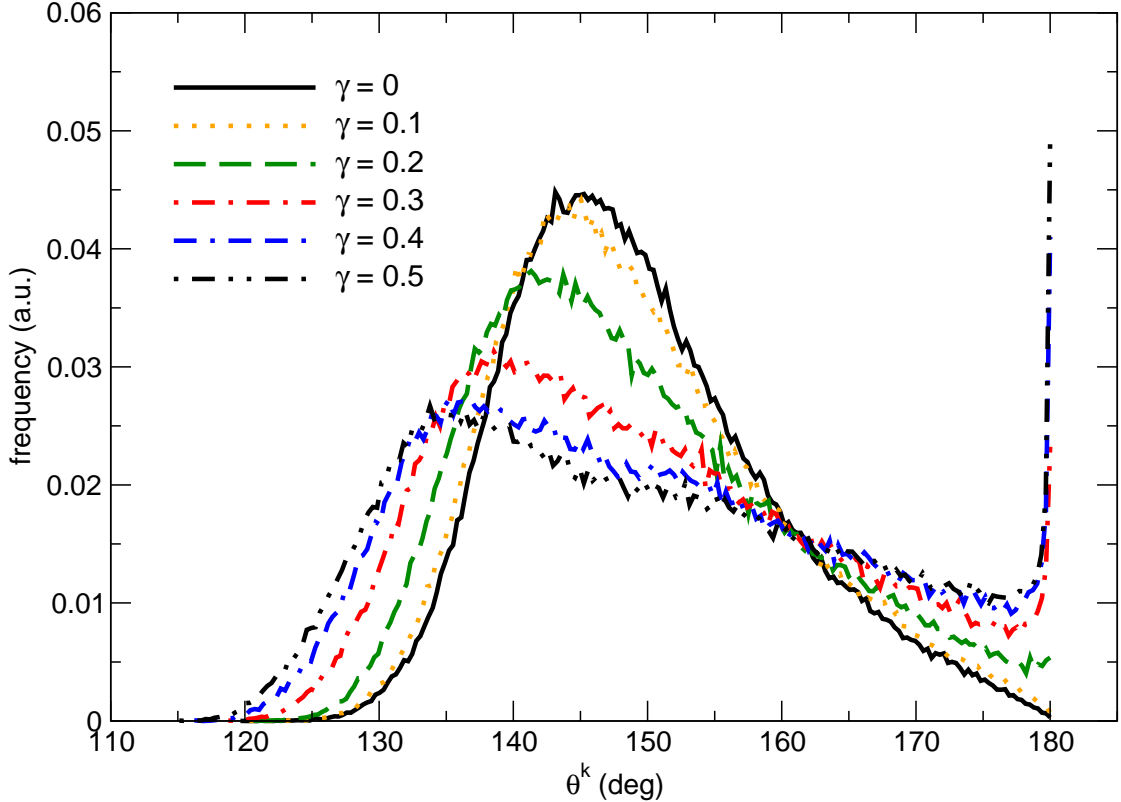}
  \caption{\label{fig:angle2} Histogram of the bond angles $\theta^k$ corresponding to shear strains ranging from $0$ to $50\%$. Starting from $\gamma=30\%$ a set of fully stretched chains appears and grows.}
\end{figure}

For the local stress tensor $\sigma^i_{\alpha\beta}$ computed as explained in Section
\ref{sec:numsim}, we define the local axis of principal tension, $t^i_\alpha$, as the normalized
eigenvector corresponding to its largest positive eigenvalue (we take $t^i_x\ge0$ to fix an
arbitrary orientation). The direction of the local tension is quantified by the angle $\phi^i$
between the projection of $t^i_\alpha$ onto the plane of shear ($xy$) and the positive $x$ axis. For
each particle $k$ having two neighbors, i.e.\ each particle that is \emph{not} a crosslink, we then
consider the angle $\theta^k$ defined by
$\arccos(\frac{\vct{r}_{ki}\cdot\vct{r}_{kj}}{r_{ki}r_{kj}})$, $i$ and $j$ being the two neighbors
of the particle, i.e.\ the angle formed by the two bonds departing from the
  particle. These angles approach $180^\circ$ in a fully stretched chain. In Figs.~\ref{fig:angle1}
and \ref{fig:angle2} we plot the histograms obtained for $\phi^i$ and $\theta^k$ over the whole gel,
at different values of the deformation.

The data show that, upon increasing the deformation, the non-linear stiffening regime corresponds
indeed to a progressive alignment of shear stresses, causing the appearance of a peak in
  the histogram shown in Fig.~\ref{fig:angle1}. {The position of the peak, i.e.\ the most likely
  orientation of the principal tension, coincides for small strain with the direction of the
  extensional axis of shear, namely $45$ degrees, and, at higher strains, deviates to slightly smaller angles (e.g.\ to about $38$ degrees at $\gamma=0.5$). This deviation can be understood by considering the effect of the deformation on an initially uniform distribution of angles $\phi^i$ at $\gamma=0$,  as explained in Appendix \ref{appendix2}.  The
  data presented in Fig.~\ref{fig:angle2} also reveal that the stress alignment corresponds indeed to the
progressive stretching of the extensible bonds in the network chains, with a set of fully stretched chains appearing and
further growing beyond $\gamma \simeq 30\%$. These results clarify the similarity with the
non-linear mechanical behaviour of biopolymer networks \citep{storm-nature2005}.} Finally, the
histograms of $\phi^{i}$ and $\theta^{k}$ values measured across the gel network indicate that the
breaking of network connections ($\gamma > 30\%$) takes place only once that the alignment of
stresses and chains has grown enough to produce fully stretched chains.

\begin{figure}
  \includegraphics[width=1.0\columnwidth, clip]{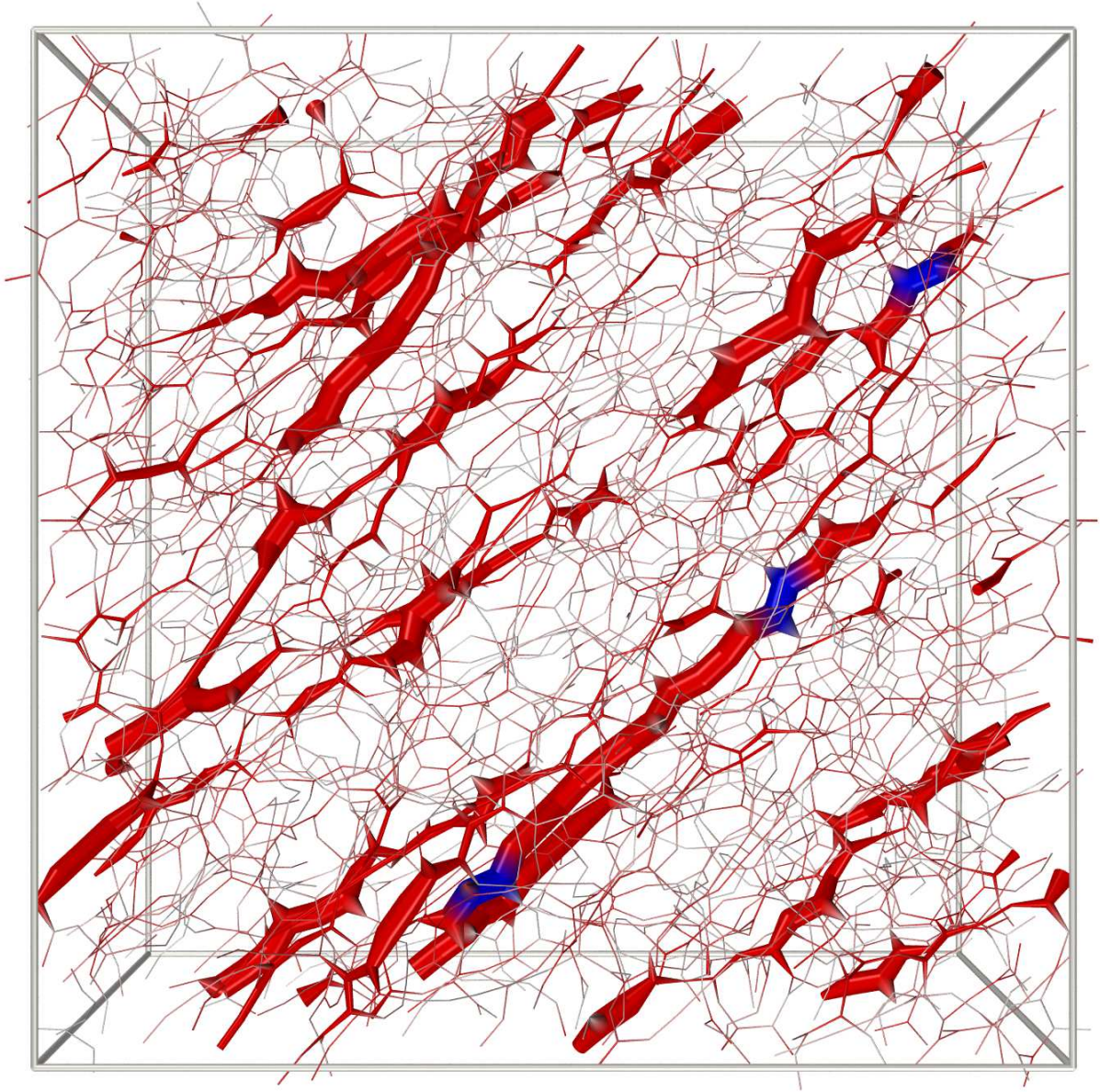}
  \caption{\label{fig:snap} A snapshot of the gel network under shear, for $\gamma = 0.35$: for clarity only the bonds are shown and the thicker strands are the ones where the tensile stress intensity is larger than $60\%$ of the maximum tensile stress. The blue indicates where breaking will happen.}
\end{figure}

By analysing the spatial distribution of stresses in the simulation snapshots at different deformation we can clearly recognise that stress intensity becomes significantly heterogeneous upon increasing the deformation beyond the linear elastic regime. The stiffening regime corresponding to the progressive stress alignment and chain stretching shown in Figs.~\ref{fig:angle1} and \ref{fig:angle2} is accompanied by a strong localization of shear stresses, whose intensity progressively concentrates on very few chains amounting to a very small part of the network structure. An example is given in Fig.~\ref{fig:snap}, that shows a snapshot of the gel network at a shear deformation of $35\%$: in the figure, bonded particles in the gel are represented by segments of thickness proportional to the intensity of local stresses. Most of the stresses produced by the deformation are in fact localized on very few chains: identifying the bonds subjected to the highest stresses allows us to predict where breaking events will occur (blue bonds in the figure). The emerging picture supports the scenario hypothesized in Section \ref{sec:loadcurve} and unravels
 the microscopic mechanisms underlying it: the stiffening of the gel is indeed due to the shear deformation pulling on the over-stretched parts of the network structure and corresponds to a progressive localization of stresses. The stress concentration eventually leads to breaking of the over-stressed bonds and triggers the yielding of the material. A previous study performed on the spontaneous relaxation dynamics of the gel has shown that tensile stresses tend to concentrate in regions of the material with higher density of crosslinks \citep{colombo_prl2013}. In agreement with that finding, here we see that bonds participating to crosslinks are the ones that preferentially break also under deformation.   

\begin{figure}
  \includegraphics[width=.9\columnwidth, clip]{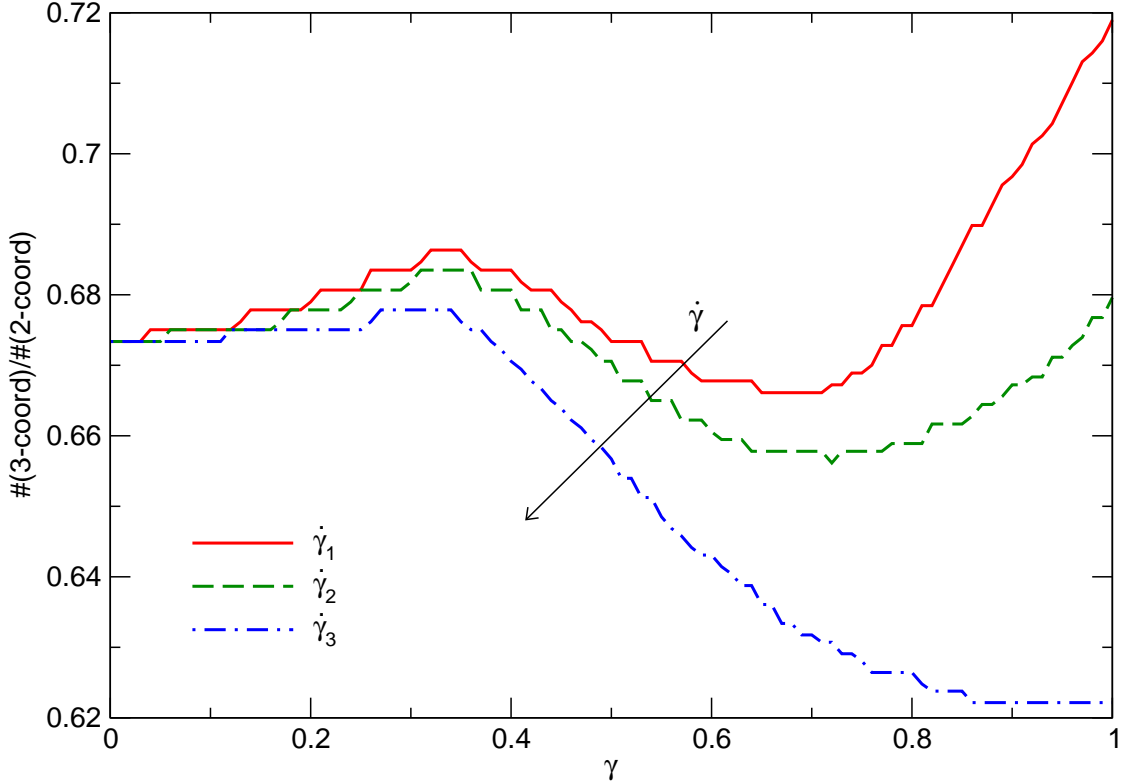}
  \caption{Number of crosslinks in the gel, expressed as a fraction of the number of two-coordinated
    particles, as a function of the amplitude of the total shear deformation $\gamma$, and for the
    three shear rates considered here. $\dot{\gamma}$ increases from top to bottom.
  \label{fig:23coordratio}}
\end{figure}

In Section \ref{sec:loadcurve}, the bond breaking and formation under increasing deformation (shown in Fig.~\ref{fig:breakbond}) suggested that yielding might be associated to significant restructuring of the gel network, because the bonds created under deformation do not necessarily contribute to bearing larger stresses in this regime.  To further investigate this point, we have analyzed in detail which type of bonds are formed under shear. At first sight, bond breaking and formation under shear may seem quite similar to the ones leading to the formation of the gel at rest and controlled by thermal fluctuations: only two- and three-coordinated particles are produced, i.e.\ the same types of connections favored by the effective interactions (see Eq.~\eqref{equ:poten}) at rest. Nevertheless, a closer look reveals important differences, as shown in Fig.~\ref{fig:23coordratio}, where we plot the fraction of crosslinks, with respect to two-coordinated particles, as deformation increases.  For the lowest shear rate considered here, the data show that after an initial slight increase, the fraction of crosslinks decreases as bond breaking proceeds, and eventually increases again for $\gamma > 70\%$. The relative amount of crosslinks attained upon yielding diminishes with increasing the shear rate and, for the highest shear rate explored, instead, it keeps decreasing to a final value that is lower than the one in the initial gel. These results, combined with the load curves obtained at different shear rates and shown in Fig. \ref{fig:breakbond}, indicate that, upon decreasing the shear rate, there is an increasing amount of excess crosslinks that are created for deformations $>70\%$. This corresponds to a decrease of the shear stresses attained in the material at the same
deformation. {Hence the crosslinks created under shear do not contribute to the stress bearing part of the material structure. This observation suggests that the onset of yielding  corresponds to a significant structural reorganisation of the gel.} 

\begin{figure}
  \includegraphics[width=.9\columnwidth, clip]{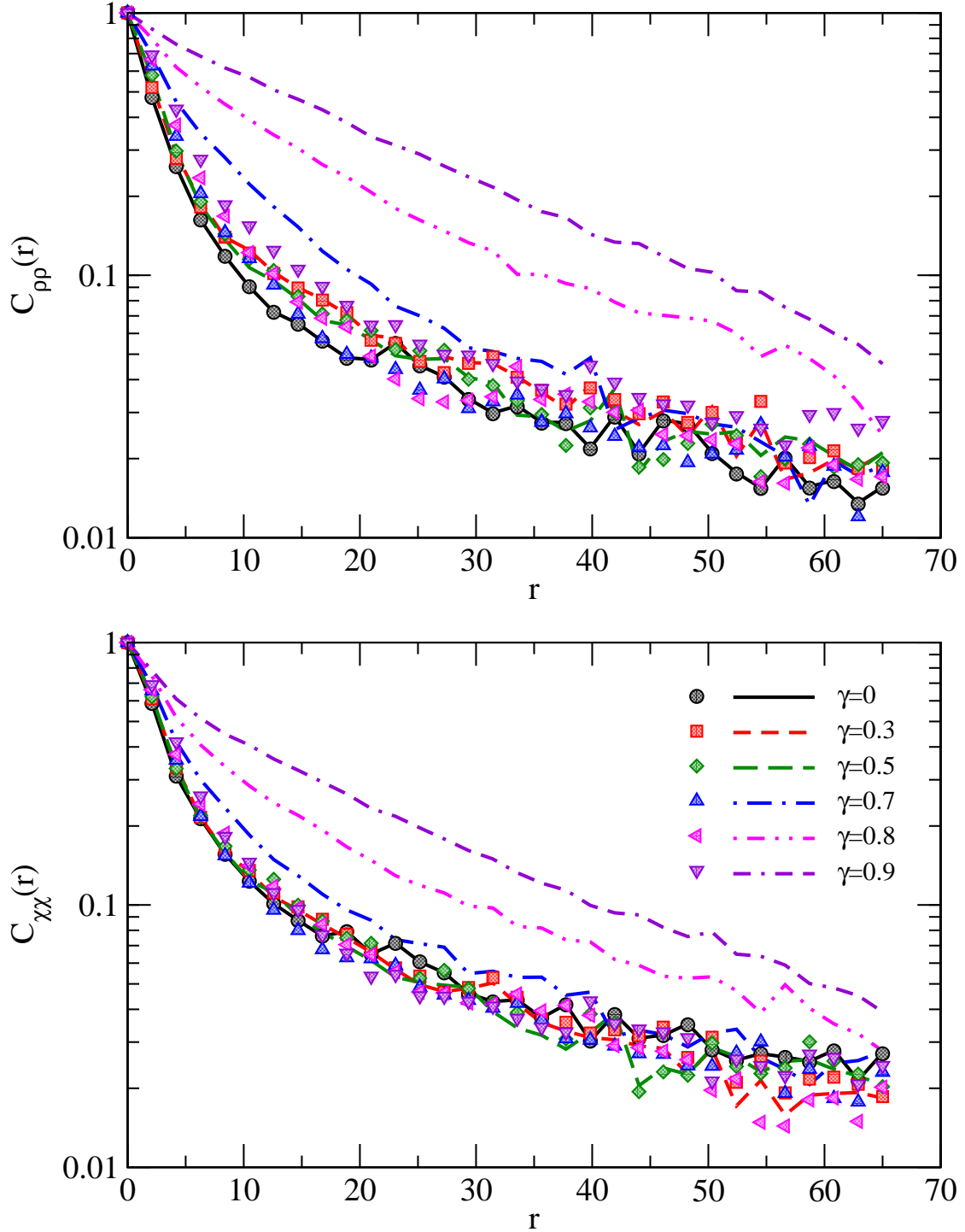}
  \caption{Spatial correlation functions of the coarse-grained density field $\rho(\vct{r})$ (top
    panel), and of the coarse-grained crosslink density field $\chi(\vct{r})$ (bottom panel); the
    coarse-graining radius $c$ equals $3\sigma$. The curves refer to various levels of strain
    $\gamma$ for a deformation at small shear rate ($\dot{\gamma}_1$, lines) and for one at high
    shear rate ($\dot{\gamma}_3$, symbols).\label{fig:ddcorr}}
\end{figure}

{To obtain a quantitative spatially resolved analysis, we
  have computed spatial correlations of local density fields at different deformation amplitudes. Using the local particle density field
\begin{equation}
  \rho(\vct{r}) = \frac{3}{4\pi c^3} \sum_{i=0}^N \Theta(c-\|\vct{r}_i-\vct{r}\|)\,,
\end{equation}
where $\Theta(\cdot)$ is the step function and $c$ is a coarse-graining radius (we omit the
dependence on $c$ for notational convenience), we compute the spatial correlation function\begin{equation}
  C_{\rho\rho}(\vct{r}) =
\frac{\left\langle\rho(\vct{s})\rho(\vct{s}+\vct{r})\right\rangle-\left\langle\rho(\vct{s}
)\right\rangle^2 } 
{\left\langle\rho^2(\vct{s})\right\rangle-\left\langle\rho(\vct{s})\right\rangle^2 }\,.
\end{equation}
With the aim of evaluating the range of spatial correlation, here for simplicity we consider its spherical average (in view of the strain induced anisotropy, one can also expand the angular part for a uniaxial symmetry around the direction of maximum extension and evaluate higher order terms that become non negligible upon increasing the deformation amplitude).
We also consider the coarse-grained \emph{crosslink} density field}
\begin{equation}
  \chi(\vct{r}) = \frac{3}{4\pi c^3} \sum_{i=0}^N \Xi_i \,\Theta(c-\|\vct{r}_i-\vct{r}\|)\,,
\end{equation}
where $\Xi_i=1$ if particle $i$ is a crosslink, $0$ otherwise, and its spatial correlation function
$C_{\chi\chi}(r)$. We plot the density correlation function $C_{\rho\rho}(r)$ in the upper panel of
Fig.~\ref{fig:ddcorr}. Here we have chosen the coarse-graining radius $c = 3\sigma$ and 
verified that this choice does not significantly affect the results. The curves refer to various
levels of strain $\gamma$ for a deformation at small shear rate ($\dot{\gamma}_1$, lines) and for
one at high shear rate ($\dot{\gamma}_3$, symbols). In the linear and stiffening regimes
($\gamma\lesssim0.5$) the correlation function changes little, meaning that no major restructuring is  taking place. On the contrary, the range of $C_{\rho\rho}(r)$ increases visibly
after yielding ($\gamma\gtrsim 0.7$) \emph{at low shear rate}, indicating the appearance of
extended domains whose density is different from the average one. A comparable restructuring does
\emph{not} happen at high shear rate: the range of the correlation function barely increases even
after yielding. The spatial correlation function of the crosslink density, depicted in the lower
panel of Fig.~\ref{fig:ddcorr}, shows the same features.

\begin{figure}
  \includegraphics[width=.9\columnwidth, clip]{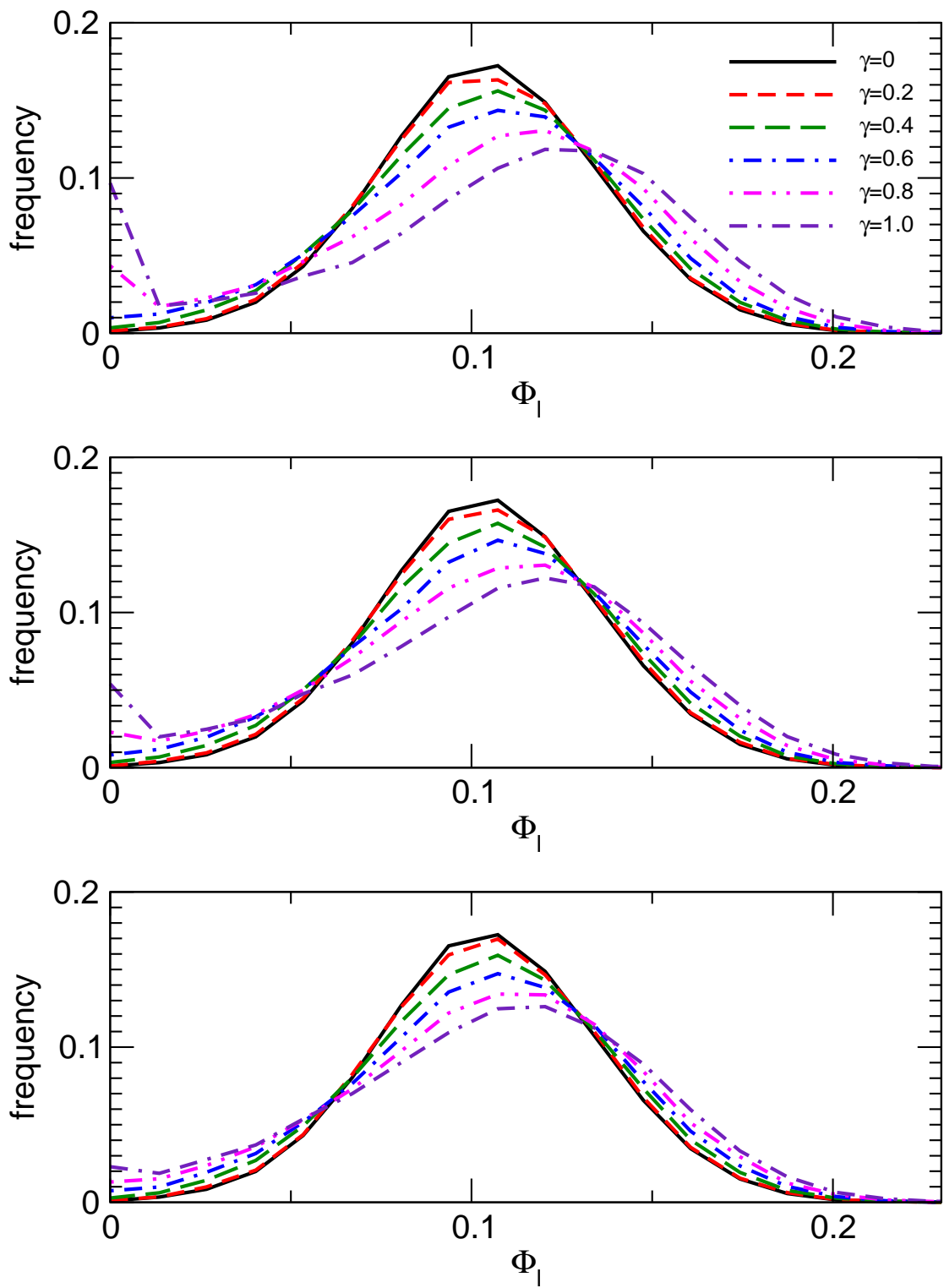}
  \caption{Local volume fraction computed in cubic boxes of linear size L/40 in the gel subjected to
    shear deformation of different amplitude $\gamma$, for the three shear rates investigated here
    ($\dot{\gamma}$ increases from top to bottom).
  \label{fig:localdens}}
\end{figure}

{To more directly unravel the restructuring in the gel upon yielding} we have monitored
the local particle volume fraction, by
dividing the simulation box in 40x40x40 sub-volumes and computing the effective volume fraction
$\Phi_{l}$ in each of them for different deformations. In Fig.~\ref{fig:localdens} we plot the
histogram of $\Phi_{l}$ values in the gel, obtained for different $\gamma$ and for the three shear
rates considered here. The histogram measured in the gel at rest is symmetric and of course centered
around the value of the total volume fraction. Upon increasing the deformation, the data becomes
asymmetric, the maximum moves to higher values of the volume fraction and indicates the presence of
extended {\it empty} regions. One can also recognize that this difference in the local volume
fraction in the yielding gel decreases significantly upon increasing the shear rate, especially the
empty regions tend to become hardly detectable.  These findings suggest that during the yielding
regime, and possibly in the non-linear regime at lower deformations, microscopic plastic
processes can take place in the gel, as well as local rearrangements that do not necessarily follow
the imposed deformation. To better elucidate this point, in the following we will quantify
non-affine displacements and irreversible processes.

\section{Non-affine displacements and residual stresses}
\label{sec:nad}
\begin{figure}
  \includegraphics[width=.9\columnwidth, clip]{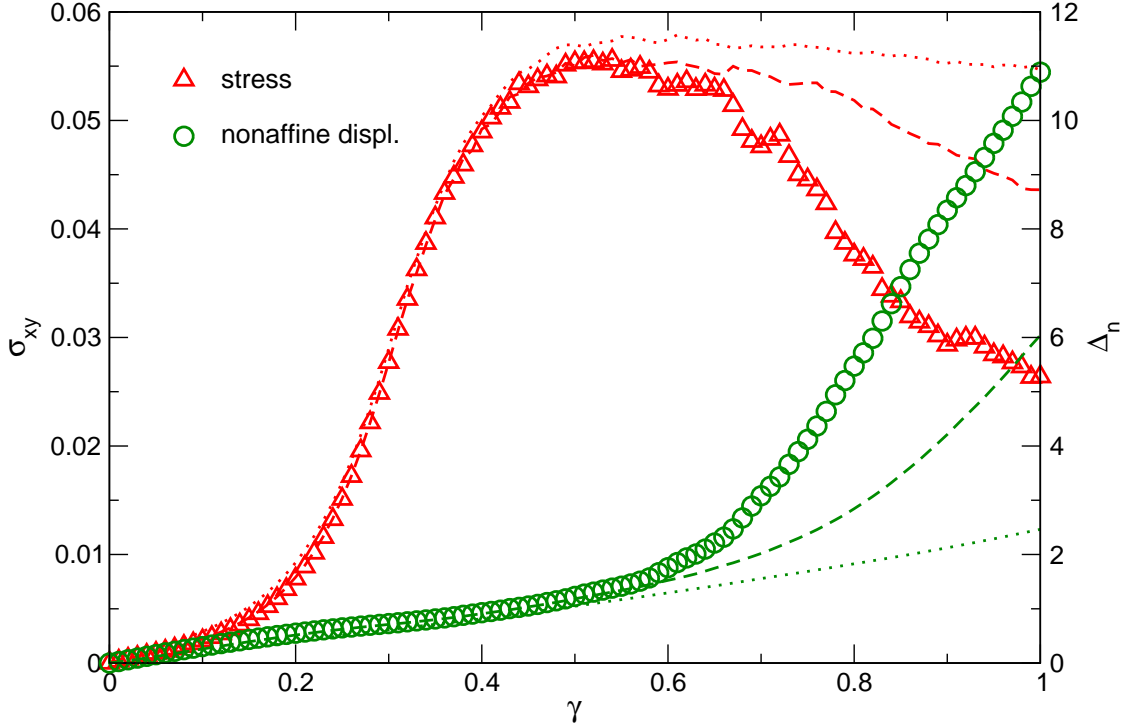}
  \caption{\label{fig:nad} Average amplitude of non-affine displacement and stresses as a function of $\gamma$. Symbols: lowest shear rate; dashed and dotted lines: intermediate and high shear rate, respectively.}
\end{figure}

We compute $\Delta_n$, the average magnitude of particle non-affine displacements with respect to the initial configuration as explained at the end of Section~\ref{sec:numsim}. In Fig.~\ref{fig:nad} we plot this quantity, averaged over all particles, as a function of $\gamma$, together with the load curve obtained at different shear rates. Interestingly, one can see that non-affine displacements are already present at low deformation, in the linear elastic regime and in the stiffening one, before any bond breaking is observed. These non-affine displacements, detected at relatively small strains and not associated to plastic processes, are apparently not affected by the shear rate and are probably due to the extended floppy modes of the gel structure \citep{alexander,rovigatti2011}, similarly, to a certain extent, to what hypothesized for semiflexible polymer gels \citep{head-pre2003,heussinger-prl2006,lieleg-prl2007}. We notice that here the increase of non-affine displacement with the strain does not seem to change between the linear elastic and the stiffening regimes.
Once parts of the gel structure start to break and the yielding sets in, instead, the magnitude of non-affine displacements dependence on the strain $\gamma$ changes qualitatively: its increase becomes much faster and depends strongly on the shear rate, with higher shear rates significantly reducing the non-affine displacements. 

\begin{figure}
  \includegraphics[width=.9\columnwidth, clip]{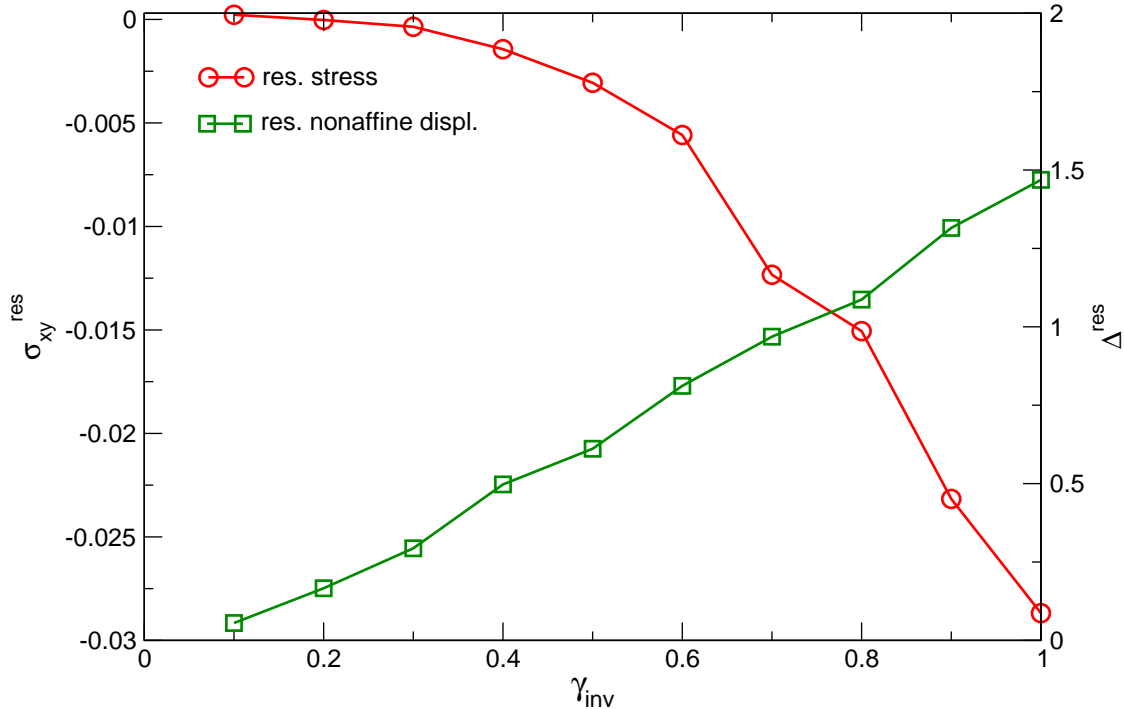}
  \caption{\label{fig:residuals} Residual stress (circles, left $y$ axis) and residual non-affine displacement (squares, right $y$ axis). The data refer to a small system (4000 particles) deformed at the low shear rate $\dot{\gamma}_1$.}
\end{figure}

{In order to investigate the connection between nonaffine displacements and plasticity, we
  use the following procedure. Starting from the undeformed gel configuration ($\gamma=0$), we
  shear the gel at the lowest shear rate $\dot{\gamma}_1$ up to a prescribed inversion strain
  $\gamma_{\rm inv}$. Then, we invert the direction of shearing and, with the same rate
  $\dot{\gamma}_1$, we bring the gel back to zero strain. At this point we measure the stress in the
  material, which we call residual stress $\sigma_{xy}^{\rm res}$, and the nonaffine displacement
  with respect to the initial configuration---the residual nonaffine displacement $\Delta^{\rm
    res}$.} In Fig.~\ref{fig:residuals} the residual stresses and the residual non-affine
displacement magnitude, accumulated up to the deformation $\gamma_{\rm inv}$, are plotted as a
function of $\gamma_{\rm inv}$. The data confirm that, up to $\gamma \simeq 30\%$, where bond
breaking under deformation starts, there is no significant macroscopic plasticity in the
material. Whereas non-affine displacements continuously increase as a function of the deformation
amplitude, once that deformation starts to induce breaking of the gel structure, the residual
stresses increase rapidly with the deformation applied. Hence the end of the stiffening regime and
the yielding correspond indeed to a non-negligible plasticity in the material response, consistent
with the shear rate dependence shown by the formation of excess crosslinks and by the change in the
local volume fraction (see Figs.~\ref{fig:23coordratio} and \ref{fig:localdens}). These findings
suggest that the plastic behavior of the material is intimately related to the mesoscale
reorganization of the gel structure under shear, that accompanies its yielding. We have seen that
this corresponds to the appearance of regions with significantly higher and significantly lower
solid volume fraction, with respect to the initial microstructure. We would like now to analyze the
effect of this micro-structural changes on the local deformations.

\section{Shear banding}
\label{sec:shearband}
The non-affine displacement plot in Fig.~\ref{fig:nad} suggests that in the yielding regime the deformation of the gel might depart significantly from a homogeneous shear flow, the more so the smaller the shear rate is. The local density histograms in Fig.~\ref{fig:localdens}, by showing that in the same regime some regions of the system densify while others are depleted of particles, point to the same fact. It is indeed well known that many soft materials (including foams, emulsions and colloidal suspensions) behaving as yield stress fluids exhibit heterogeneous yielding dynamics~\citep{vermant_jor2009,ovarlez-rheoacta2009,fall-prl2010,rajaram_sm2010,gibaud2010heterogeneous}. These often take the form of \emph{shear bands}, regions of unequal shear rate coexisting in the material at a common value of the shear stress~\citep{schall2009shear, moller2008shear}; they can be either transient, marking the transition to a homogeneous flow, or permanent~\citep{martin2012transient,fielding-arxiv2013}.

In order to investigate whether the yielding of our model system might be associated to such mechanical inhomogeneities, we compute during each strain step a displacement profile, by evaluating the average particle displacement in the direction of shear ($x$) as a function of the gradient direction ($y$). More precisely, we divide the simulation box in slabs of width $\Delta=L/100$, $L$ being the box size, along the gradient coordinate ($y$) and compute in each slab the $x$ component of the average particle displacement occurring in the $n$-th strain step:
\begin{equation}
  d_n(y) = \frac{1}{M}\sum_{|y_{i,n}-y|<\Delta/2}(x_{i,n}-x_{i,n-1})
\end{equation} 
where the summation extends to all particles that at the end of the step are found inside the slab, and $M$ represents their number.

\begin{figure}
  \includegraphics[width=.9\columnwidth, clip]{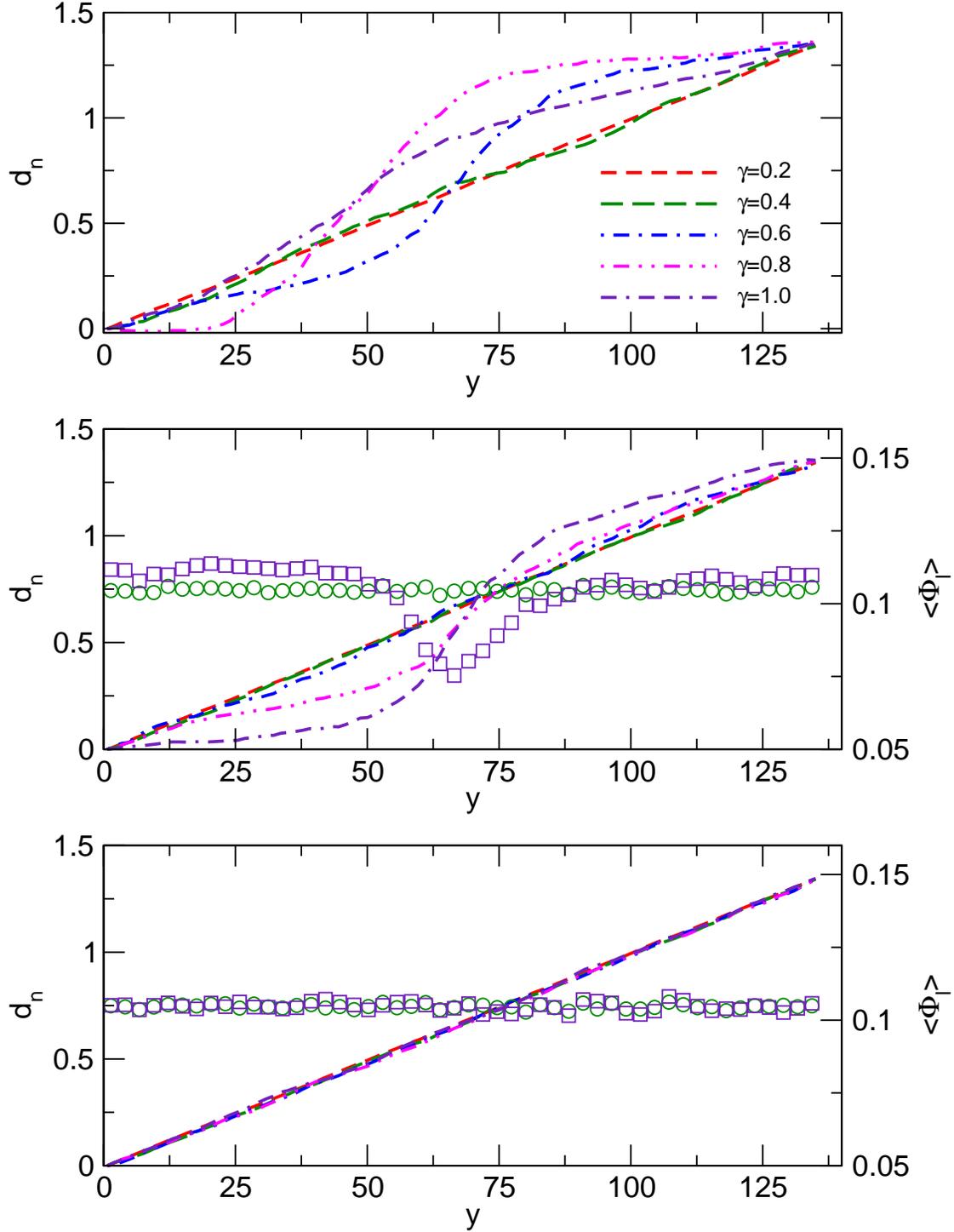}
  \caption{\label{fig:velprof} Displacement profiles at selected values of the cumulative strain $\gamma$ for the three different shear rates, increasing from top ($\dot{\gamma}_1$, low) to bottom ($\dot{\gamma}_3$, high). The displacements are relative to the displacement of the bottom layer, so that $d_n(y=0)=0$ for all curves. In the second and third panels the open symbols, referring to the right $y$ axis, represent the density profile at $\gamma=0.4$ (circles) and $\gamma=1.0$ (squares), respectively.}
\end{figure}

The displacement profile $d_n(y)$ may be interpreted as a velocity profile upon dividing by the duration $\delta t$ of the elementary strain step. We report in Fig.~\ref{fig:velprof} the
displacement profiles for selected values of the cumulative shear strain $\gamma$; the three panels refer to different values of the strain rate, from low ($\dot{\gamma}_1$, top) to high ($\dot{\gamma}_3$, bottom). For ease of visualization and comparison we plot displacements relatively to the displacement of the bottom layer, so that $d_n(y=0)$ equals zero for all curves.
Before any yielding occurs ($\gamma\lesssim0.4$) the profiles are linear regardless of the strain rate, meaning that the deformation is macroscopically homogeneous. Of course even in this regime there exist microscopic displacements that are non-affine; nonetheless, the average deformation in the direction of shear follows the overall imposed gradient without any macroscopic inhomogeneity (i.e.\ inhomogeneity comparable in size to the system size). In the case of the high strain rate ($\dot{\gamma}_3$, bottom panel) this is so even in the yielding regime, presumably because the gel does not have time to significantly rearrange between one strain step and another and major inhomogeneities cannot develop. On the contrary, in the case of the lower strain rates the average deformation profiles develop, upon yielding, inhomogeneities on length scales comparable to the system size, i.e.\ shear bands. Because of the periodic boundary conditions used in the simulations, the profiles are bound to have the same slope at the bottom and at the top of the box, but, apart from this, they are reminiscent of the ones typically measured in experiments~\citep{divoux2012yielding}. It can also be observed that, in the case of the lowest shear rate ($\dot{\gamma}_1$, top panel) the bands develop earlier ($\gamma\approx0.6$) than in the case of the intermediate rate ($\dot{\gamma}_2$, middle panel, where bands appear at $\gamma\approx0.8$). This finding seems to fit in with recent experimental investigations in systems displaying transient shear banding, in which the authors found that the duration of the transient bands is inversely proportional to the magnitude of the strain rate~\citep{divoux2010transient}. It is also consistent with the results of numerical simulations in dense jammed suspensions in \citet{chaudhuri-pre2012}.

\begin{figure}
  \includegraphics[width=\columnwidth]{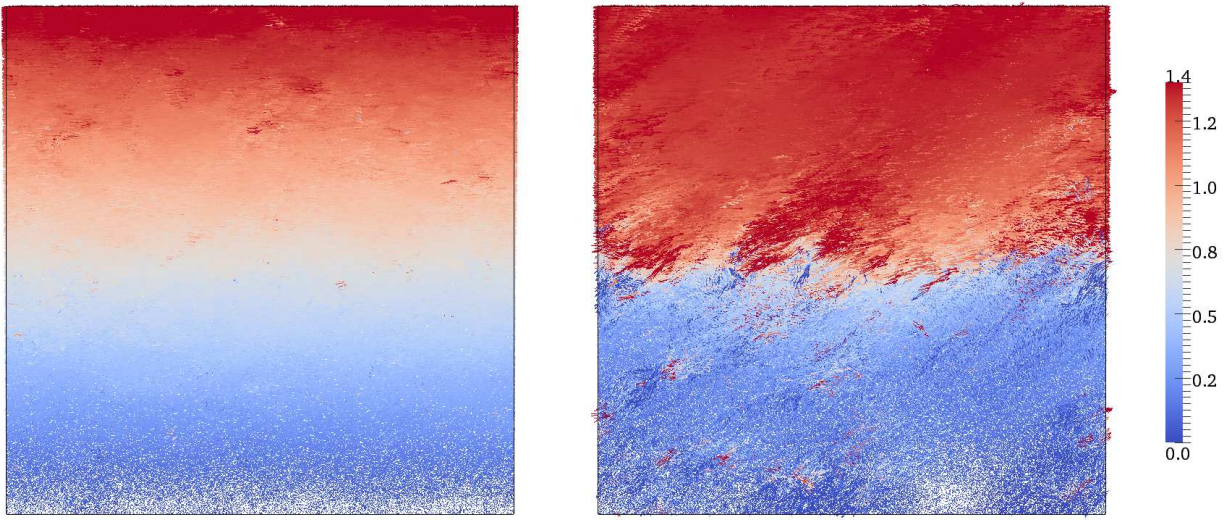}
  \caption{\label{fig:shearband} Spatial map of particle displacements occurring in a elementary strain step at $\gamma=0.1$ (left) and $\gamma=0.6$ (right) for the lowest shear rate ($\dot{\gamma}_1$). The color code represents the component of the displacement in the direction of shear.}
\end{figure}

In Fig.~\ref{fig:shearband} we present spatial maps of particle displacements, providing additional evidence for the existence of shear bands. We project all the particles onto the plane of shearing ($xy$) and represent the displacement of each particle in a given strain step as an arrow; the color of the arrow varies with the $x$ component of the displacement, from small (blue) to large (red). The two maps refer to different values of the cumulative strain: $\gamma=0.10$ in the left map, whereas $\gamma=0.60$ in the right map; in both cases the gel is being sheared at the lowest strain rate ($\dot{\gamma}_1$). While in the first case the displacement field increases at constant rate from bottom to top, in the second case the rate of variation in the middle of the box is much faster, giving rise to a shear band.

We end this section by discussing only briefly the microscopic origin of shear banding in our model gel. A thorough analysis of this phenomenon requires a deeper investigation that is out of the scope of this paper. Future work will be devoted, for example, also to analyze whether in our model system shear bands are transient, eventually leading to a homogeneous flow profile, or permanent, in the context of the experiments~\citep{divoux2010transient}.
Figure~\ref{fig:localdens} suggests that in the present system shear banding couples to the fluctuations in the local density of particles. Figure~\ref{fig:velprof} shows that this is indeed the case: the open symbols appearing in the second and third panels show the gel density profile along the gradient direction obtained in the same way as the displacement profile. It is apparent that shear banding correlates with inhomogeneities in the density profile: the layers of lower density are also the layers where the higher displacement gradients appear. The emerging scenario is that, beyond the stiffening regime and as yielding sets in, deformation induces density fluctuations that are absent at rest and gradually lead to the formation of colloid-rich and colloid-poor regions in the gel. Shear induced bond formation results into an excess of crosslinks in denser regions, that are therefore presumably stiffer. Nevertheless, these localized excess crosslinks do not necessarily reinforce the stress-bearing backbone of the gel. On the contrary, the presence of extended colloid-poor domains leads to an overall lower average stress and probably helps developing a slip plane (see Fig.~\ref{fig:shearband}) and the shear banding. Another possibility is instead that the shear banding develops earlier and induces  the fluctuations in the local density of the gel. Elucidating this point calls for further work, but it seems  very likely that  the two phenomena are strongly coupled and reinforce each other.

{Another important point is the role of hydrodynamics, which is not included in our simulations. Because the phenomena we are focusing on develop upon lowering the shear rate and the modified P\'eclet number $\tilde{Pe} \ll 1$, we expect that the scenario described here is not qualitatively modified by the presence of hydrodynamic interactions. Nevertheless, they could well play a role in the major reorganisation of the gel upon yielding, especially for higher shear rates \citep{Bergenholtz-jfm2002}, hence including hydrodynamics in this type of study, although still challenging, would be extremely interesting.}

\section{Conclusions}
We have investigated the mechanical response of a model colloidal gel under shear, with a new numerical procedure that allows to follow deformations at relatively low shear rates. Moreover, differently from most existing numerical studies, we do not impose an artificial gel structure nor specific microscopic processes: the gel network and its mechanical response are the result of the interplay between the same effective interactions that stabilize the gel structure at rest and the imposed shear deformation. The effective interactions include a bending rigidity for the colloidal bonds and lead to the formation of a soft solid at relatively low solid volume fractions in a colloidal suspension, consistent with experimental observation. The load curve of the gel displays, after an initial linear elastic response for deformation amplitudes $\gamma \le 10\%$, a significant stiffening and, at 
$\gamma \simeq 50\%$, a final yielding. By means of a space-resolved analysis of stresses and deformations in the gel network, we have shown that the strain hardening corresponds first to a progressive alignment of stresses and stretching of parts of the structure (the {\it chains} or less connected parts of the gel network) along the direction of maximum extension under shear, and eventually to the shear deformations pulling on the over-stretched chains. This phenomenon is qualitatively similar to the strain stiffening typical of biopolymer networks.  We also find that the strain hardening of the gel is accompanied by an increasing stress localization, with the highest stress magnitude concentrating in few parts of the structure, where eventually, starting from $\gamma \simeq 30\%$, breaking occurs. Because of the strong localization of stresses in the material, a small fraction of broken bonds is sufficient to trigger the yielding. Once yielding sets in, we observe the arising of shear induced density fluctuations in the gel, that seems to {\it coarsen} into colloid-rich and colloid-poor domains. This phenomenon is apparently coupled to the onset of significant strain inhomogeneities in the material that can be related to a shear banding, very similar to what observed in recent experiments. We have been able to elucidate here that the strain-induced reorganization of the gel and the shear banding are accompanied by an excess of crosslinks that are formed preferentially in the denser regions and therefore increase only the local stiffness of those domains, but not the overall capacity of the yielded gel to support higher stresses. This phenomenon is reminiscent of shear induced structural changes in other gels, that might arise from the imposed deformations acting differently on softer and stiffer domains \citep{onuki}, and it might also have some similarities with the coupling between shear deformations and density fluctuations proposed for complex fluids \citep{milner-pre93,schmitt-pre95,fielding-prl2003} or with the rupture of the material \citep{Mohraz-jor2005}. The shear induced density fluctuations, the shear banding and the final level of stress attained in the gel upon yielding are strongly affected by the shear rate. Higher shear rates do not allow for rearrangements of the gel that can accommodate local changes in density, as signaled by a decreasing amount of non-affine displacements: this limits density fluctuations and the shear induced bond formation, resulting in higher stresses attained upon yielding.
  
Overall this study has offered new, significant insight into the behavior, under deformation, of soft solids and in particular of colloidal gel networks. On the one hand, the results discussed here give, for the first time, a microscopic physical picture of how phenomena like stiffening, yielding and shear banding may occur in these materials. On the other, they highlight issues such as the stress localization or the feedback between density fluctuations and shear banding or strain inhomogeneities that have been hardly addressed in numerical studies in spite of being probably ubiquitous in experimental observations. 
Our numerical investigation and methodology have therefore interesting potential to complement experiments and lead to new theoretical development in this area of research.

\section{Acknowledgments}
The authors thank for insightful discussions Daniel Blair, Thibaut Divoux, Sudeep Dutta, David Head, S\'ebastien Manneville, Peter Olmsted and Dimitris Vlassopoulos. 
This work was supported by Swiss National Science Foundation (SNSF) (Grants No. PP00P2\_126483/1).

\appendix
\section{Oscillatory rheology}
\label{appendix1}
\begin{figure}[h]
\begin{center}
  \includegraphics[width=.65\textwidth]{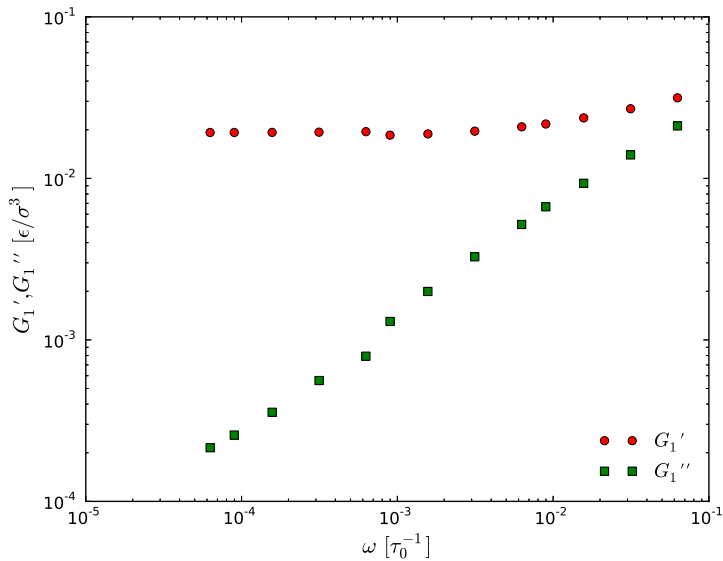}
  \caption{Oscillatory shear simulation with $\gamma_{0}=1\%$: frequency sweeps.
First-harmonic storage and loss moduli, $G'_1$ and $G''_1$, as a function of the frequency $\omega$. 
\label{fig:G0}}
\end{center}
\end{figure}

\begin{figure}[h!]
\begin{center}
  \includegraphics[width=.68\columnwidth]{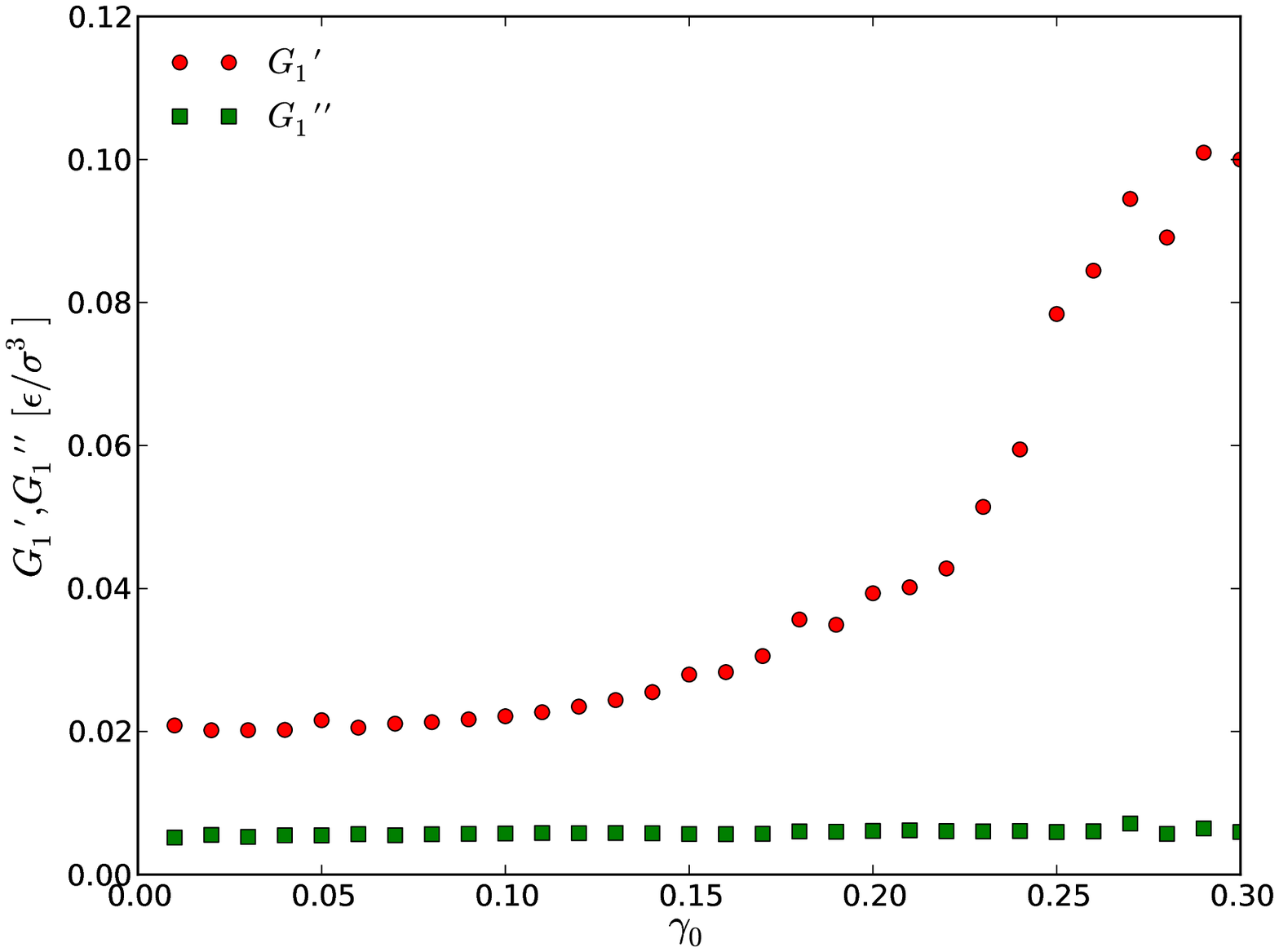}
  \caption{Oscillatory shear simulation ($\omega\approx3\cdot10^{-3}\tau_0^{-1}$): strain sweep.
First-harmonic storage and loss moduli, $G'_1$ and $G''_1$, as a function of the strain amplitude
$\gamma_0$.\label{fig:G1}}
\end{center}
\end{figure}

\begin{figure}[h]
\begin{center}
  \includegraphics[width=.8\columnwidth]{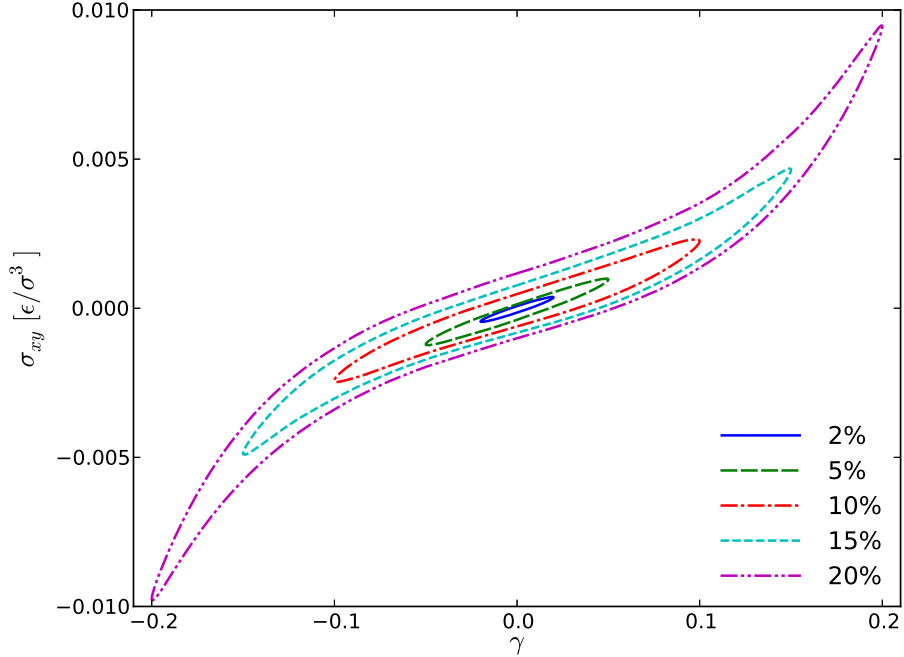}
  \caption{Oscillatory shear simulation ($\omega\approx3\cdot10^{-3}\tau_0^{-1}$): Lissajous-Bowditch
plot. The different curves refer to different values of the strain amplitude $\gamma_0$.
\label{fig:lissajous}}
\end{center}
\end{figure}

We have performed a limited set of simulations under oscillatory conditions on a gel sample with
4000 particles. The shear strain $\gamma$ was modulated periodically according to the equation
$\gamma(t)=\gamma_0\sin(\omega t)$, with a fixed angular frequency
$\omega\approx3\cdot10^{-3}\,\tau_0^{-1}$ and a strain amplitude $\gamma_0$ increasing from 1\% to
30\% in 1\% steps. The value of the angular frequency was chosen such that the maximum rate of
deformation $\omega\gamma_0$ falls within the range of shear rates ($\sim
10^{-5}\div10^{-3}\,\tau_0^{-1}$) investigated for the steady shear startup in the main text. The
discrete-step deformation protocol that we have used for the steady shear simulations is not readily
applicable to an oscillatory setup. We have instead used for this purpose the following equations of
motion:
\begin{equation}
m\frac{d^2\vec{r}_i}{dt^2} =
-\xi\left(\frac{d\vec{r}_i}{dt}-\dot{\gamma}(t)y_i\vec{e}_x\right)-\nabla_{\vec{r}_i}U\,,
\end{equation}
where $\vec{e}_x$ is a unit vector along the $x$ axis. For the integration of the equations of
motion we have used the same timestep as in the steady shear simulations; the shear strain $\gamma$
and the Lees-Edwards boundary conditions were updated at each simulation step. By monitoring the
time evolution of the shear stress $\sigma_{xy}(t)$ one can extract the $n$-th order viscoelastic
coefficients $G'_n$ and $G''_n$, $n\ge1$, in the following way:
\begin{align}
G'_n(\omega,\gamma_0) &= \frac{\omega}{\gamma_0\pi}\int_{t_0}^{t_0+2\pi/\omega}\sigma_{xy}(t)\sin(n\omega
t)\, dt\,, 
\\ G''_n(\omega,\gamma_0) &=
\frac{\omega}{\gamma_0\pi}\int_{t_0}^{t_0+2\pi/\omega}\sigma_{xy}(t)\cos(n\omega t)\, dt\,;
\end{align}
at steady state the result is independent of the choice of the time instant $t_0$. In Fig.~\ref{fig:G0} we show the results of the frequency sweeps performed with $\gamma_{0}=1\%$, in the linear response regime, showing that $G''_1$ increases linearly with $\omega$ and the values of $G'_1$ are consistent with the ones measured in the steady shear protocol. We plot in Fig.~\ref{fig:G1} the first-order storage and loss moduli, $G_1'$ and $G_1''$, as a
function of the strain amplitude $\gamma_0$. An inspection of the plot confirms the set of features
that we could already identify from the stress-strain curves in steady shear conditions. The gel
behaves as a solid material ($G_1'\gg G_1''$), with a storage modulus that is roughly constant
($\approx 2\cdot10^{-3}\epsilon/\sigma^3$) in the range $\gamma_0\lesssim0.1$. 
\begin{figure}[h!]
\begin{center}
  \includegraphics[width=.68\columnwidth]{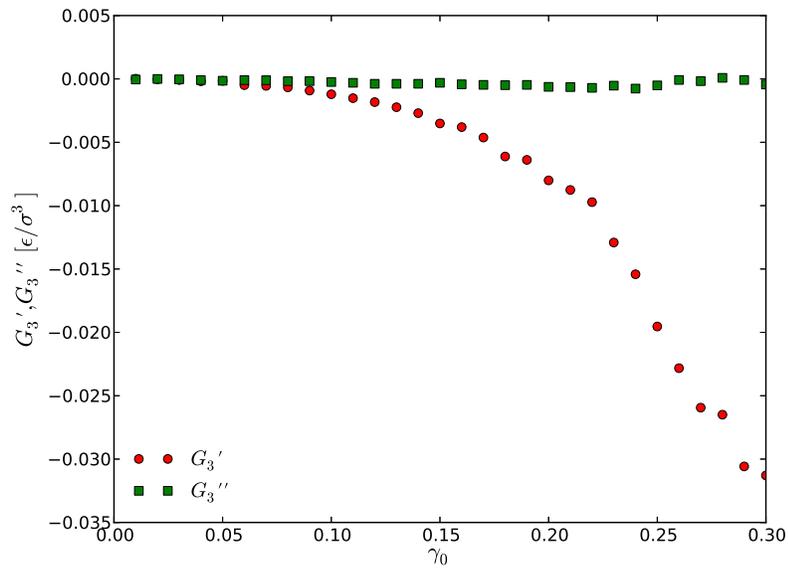}
  \caption{Oscillatory shear simulation ($\omega\approx3\cdot10^{-3}\tau_0^{-1}$): strain
sweep. Third-harmonic storage and loss moduli, $G'_3$ and $G''_3$, as a function of the strain amplitude
$\gamma_0$.\label{fig:G3}}
\end{center}
\end{figure}
After this initial
linear regime, the gel undergoes a pronounced stiffening: the storage modulus $G_1'$ increases with
the strain amplitude $\gamma_0$, whereas the loss modulus $G_1''$ is constant.
To further characterize the nonlinear regime we have also calculated the third-order viscoelastic
coefficients $G_3'$ and $G_3''$, which we plot in Fig.~\ref{fig:G3}. The third-order coefficient
$G_3'$ becomes progressively more negative, a behavior that characterizes strain-stiffening
materials (see for instance Ewoldt, Hosoi and McKinley, J. Rheol. 52, 1427 (2008)), whereas the nonlinear loss modulus $G_3''$ is essentially
zero.
Finally, we present in Fig.~\ref{fig:lissajous} a set of Lissajous-Bowditch plots, i.e. plots
of shear stress vs. shear strain under oscillatory conditions, for different values of the strain
amplitude $\gamma_0$. While for small $\gamma_0$ the curves are ellipses, signaling a linear viscoelastic
response, for larger deformations intra-cycle strain stiffening is clearly visible, without any
apparent increase in the dissipation.
The physical picture for the nonlinear regime that we get from these results is completely
consistent with the one resulting from the steady shear startup simulations reported in the main
text.

\section{Alignment of the interparticle bonds with increasing strain}
\label{appendix2}

The most likely orientation of the principal tension shown in Fig.~\ref{fig:angle1} begins at
$45^\circ$ (i.e.\ the direction of the extensional axis of shear), but systematically deviates to
smaller angles with increasing strain (e.g.\ to about $38^\circ$ at $\gamma=0.5$). We can rationalize this observation as follows.

\begin{figure}
\begin{center}
  \includegraphics[width=.7\columnwidth]{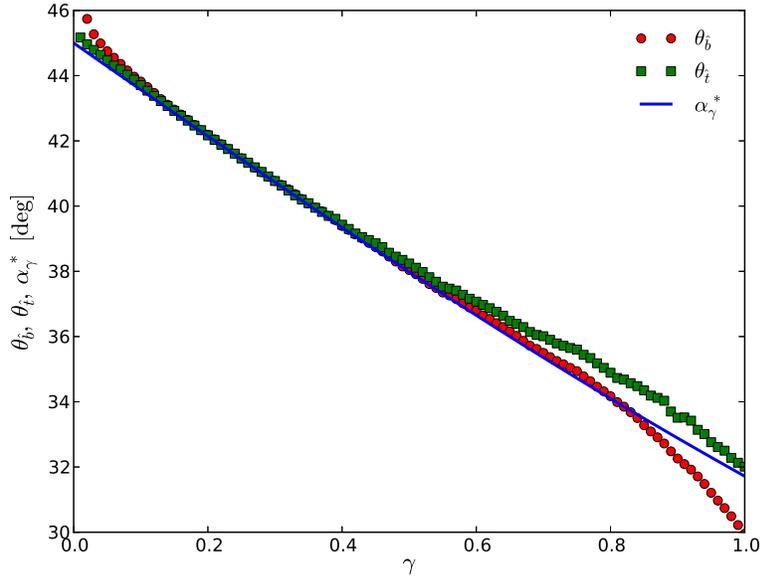}
  \caption{Most likely bond orientation $\theta_{\hat{b}}$, direction of principal tension
    $\theta_{\hat{t}}$, and outcome of the affine deformation model $\alpha_\gamma^*$, as a function
    of the shear strain $\gamma$.\label{fig:align}}
\end{center}
\end{figure}

\emph{(a) The most likely stress orientation follows the most likely bond orientation.} We already
know from the stress decomposition shown in Fig.~\ref{fig:loadcurve} that an important part of the
stress---most of it in the stiffening regime---is borne by the stretching of the interparticle
bonds. It is therefore reasonable to expect that the direction along which the material can sustain
the largest tension correspond to the most likely orientation of the interparticle bonds.  We define
the most likely bond orientation $\hat{b}$ as in the following:
\begin{equation}
  \hat{b}=\operatorname*{arg\,max}_{\vec{v}\,,\|\vec{v}\|=1} \sum_k
\left(\vec{b}_k\cdot\vec{v}\right)^2\,,
\end{equation}
where $k$ runs over the set of the interparticle bonds. Similarly, as already stated in
Sec.~\ref{sec:spaceresolved} we define the direction of principal tension $\hat{t}$ as the
normalized eigenvector corresponding to the largest (positive) eigenvalue of the global stress
tensor $\sigma_{\alpha\beta}$. We further denote by $\theta_{\hat{b}}$ (resp.  $\theta_{\hat{t}}$)
the angle that the projection onto the $xy$ plane of the vector $\hat{b}$ (resp.  $\hat{t}$) forms
with the positive $x$ axis; we take $\hat{b}_x\ge0$ (resp. $\hat{t}_x\ge0$) to fix an orientation.
In Fig.~\ref{fig:align} we plot $\theta_{\hat{b}}$ and $\theta_{\hat{t}}$ as a function of the shear
strain $\gamma$ for the lowest shear rate $\dot{\gamma}_1$: the two quantities are indeed close to
each other. Note that the two match perfectly in the stiffening regime, confirming that bond
stretching is the mechanically most relevant process to be considered there.

\emph{(b) The most likely bond orientation begins at 45$^\circ$ for small strain, but deviates to
  lower angles with increasing strain.} This feature has a purely geometrical origin and can be
explained by a simple argument. Consider an interparticle bond whose projection onto the $xy$ plane
forms in the undeformed gel configuration ($\gamma=0$) an angle $\alpha_0\in[-\pi/2,\pi/2]$ with the
positive $x$ axis (we fix again an orientation by requiring that the $x$ component of the bond
vector be positive). After a \emph{purely affine} deformation the same bond vector will be
characterized by a new angle $\alpha_\gamma$ given by
\begin{equation}
  \alpha_\gamma = f_\gamma(\alpha_0) = \arctan\left(\frac{\sin
\alpha_0}{\cos\alpha_0+\gamma\sin\alpha_0}\right)\,,
\end{equation}
where $\gamma$ is the shear strain. If $\mathscr{P}_0(\alpha_0)$ is the distribution of bond angles in the
undeformed configuration, the distribution of angles after the deformation will be
\begin{equation}
  \mathscr{P}_\gamma(\alpha_\gamma) =
\mathscr{P}_0(\alpha_0)\left(\frac{df_\gamma(\alpha_0)}{d\alpha_0}\right)^{-1} = 
\mathscr{P}_0(\alpha_0)\left(1+\gamma^2\sin^2\alpha_0+2\gamma\sin\alpha_0\cos\alpha_0\right)\,.
\end{equation}
If we now assume that the initial distribution is uniform ($\mathscr{P}_0$ constant), by maximizing the
right hand side of the previous equation with respect to $\alpha_0$ we find that the angle distribution in the
deformed configuration will be peaked around the value $\alpha_\gamma^*$ given by
\begin{equation}
\alpha_\gamma^* = f_\gamma\left(\frac{\pi}{4}+\arctan\left(\frac{\gamma}{2}\right)\right)\,.
\end{equation}
This function starts at $\pi/4$ for $\gamma\to0$, but systematically deviates to smaller angles with
increasing shear strain (it approaches 0 for $\gamma\to\infty$). The prediction $\alpha_\gamma^*$ (expressed
in degrees) is represented in Fig.~\ref{fig:align} by the full line: even though the real deformation is not
completely affine, the outcome of this simple argument explains the data quite well, especially in the
stiffening regime.


\begin{thebibliography}{87}
\newcommand{\enquote}[1]{``#1''}
\providecommand{\natexlab}[1]{#1}

\bibitem[{Alexander(1998)}]{alexander}
Alexander, S., \enquote{Amorphous solids: their structure, lattice dynamics and
  elasticity,} Physics Reports \textbf{296}, 65 (1998).

\bibitem[{Arevalo \emph{et~al.}(2010)Arevalo, Urbach, and
  Blair}]{blair-collagen2010}
Arevalo, R.~C., J.~S. Urbach, and D.~L. Blair, \enquote{Size-dependent rheology
  of type-I collagen networks,} Biophysical journal \textbf{99}(8), L65--L67
  (2010).

\bibitem[{Barrat and Lemaitre(2011)}]{barrat-lemaitre}
Barrat, J.-L., and A.~Lemaitre, \emph{Heterogeneities in amorphous systems
  under shear in Dynamical Heterogeneities in Glasses, Colloids, and Granular
  Media - International Series of Monographs on Physics, edited by L. Berthier,
  G. Biroli, J.-P. Bouchaud, L. Cipelletti and W. van Saarloos} (Oxford
  University Press, 2011).

\bibitem[{Bergenholtz \emph{et~al.}(2002)Bergenholtz, Brady, and
  Vicic}]{Bergenholtz-jfm2002}
Bergenholtz, J., J.~F. Brady, and M.~Vicic, \enquote{The non-newtonian rheology
  of dilute colloidal suspensions,} Journal of Fluid Mechanics \textbf{456},
  239--275 (2002).

\bibitem[{Bianchi \emph{et~al.}(2011)Bianchi, Blaak, and Likos}]{bianchi}
Bianchi, E., R.~Blaak, and C.~N. Likos, \enquote{Patchy colloids: state of the
  art and perspectives,} Phys. Chem. Chem. Phys. \textbf{13}, 6397--6410
  (2011).

\bibitem[{Carpineti and Giglio(1992)}]{Carpineti-prl1992}
Carpineti, M., and M.~Giglio, \enquote{Spinodal-type dynamics in fractal
  aggregation of colloidal clusters,} Phys. Rev. Lett. \textbf{68}, 3327--3330
  (1992).

\bibitem[{Charbonneau and Reichman(2007)}]{charbonneu2007_pre}
Charbonneau, P., and D.~R. Reichman, \enquote{Phase behavior and
  far-from-equilibrium gelation in charged attractive colloids,} Phys. Rev. E
  \textbf{75}, 050401 (2007).

\bibitem[{Chaudhuri \emph{et~al.}(2012)Chaudhuri, Berthier, and
  Bocquet}]{chaudhuri-pre2012}
Chaudhuri, P., L.~Berthier, and L.~Bocquet, \enquote{Inhomogeneous shear flows
  in soft jammed materials with tunable attractive forces,} Phys. Rev. E
  \textbf{85}, 021503 (2012).

\bibitem[{Colombo and Del~Gado(2014)}]{colombo-sm2014}
Colombo, J., and E.~Del~Gado, \enquote{Self-assembly and cooperative dynamics
  of a model colloidal gel network,} Soft Matter \textbf{10}, 4003--4015 (2014).

\bibitem[{Colombo \emph{et~al.}(2013)Colombo, Widmer-Cooper, and
  Del~Gado}]{colombo_prl2013}
Colombo, J., A.~Widmer-Cooper, and E.~Del~Gado, \enquote{Microscopic picture of
  cooperative processes in restructuring gel networks,} Phys. Rev. Lett.
  \textbf{110}, 198301 (2013).

\bibitem[{Conrad \emph{et~al.}(2010)Conrad, Wyss, Trappe, Manley, Miyazaki,
  Kaufman, Schofield, Reichman, and Weitz}]{conrad-jor2010}
Conrad, J., H.~Wyss, V.~Trappe, S.~Manley, K.~Miyazaki, L.~Kaufman,
  A.~Schofield, D.~Reichman, and D.~Weitz, \enquote{Arrested fluid-fluid phase
  separation in depletion systems: Implications of the characteristic length on
  gel formation and rheology,} Journal of Rheology (1978-present)
  \textbf{54}(2), 421--438 (2010).

\bibitem[{Cordier \emph{et~al.}(2008)Cordier, Tournilhac, Soulie-Ziakovic, and
  Leibler}]{leibler_nature}
Cordier, P., F.~Tournilhac, C.~Soulie-Ziakovic, and L.~Leibler,
  \enquote{Self-healing and thermoreversible rubber from supramolecular
  assembly,} Nature \textbf{451}, 977 (2008).

\bibitem[{Del~Gado and Kob(2005)}]{edgkob_epl05}
Del~Gado, E., and W.~Kob, \enquote{Structure and relaxation dynamics of a
  colloidal gel,} Europhys. Lett. \textbf{72}, 1032 (2005).

\bibitem[{Del~Gado and Kob(2010)}]{delgado2010microscopic}
Del~Gado, E., and W.~Kob, \enquote{A microscopic model for colloidal gels with
  directional effective interactions: network induced glassy dynamics,} Soft
  Matter \textbf{6}(7), 1547--1558 (2010).

\bibitem[{Denn and Bonn(2011)}]{Denn-rheoacta2011}
Denn, M., and D.~Bonn, \enquote{Issues in the flow of yield-stress liquids,}
  Rheologica Acta \textbf{50}(4), 307--315 (2011).

\bibitem[{Di~Michele \emph{et~al.}(2013)Di~Michele, Varrato, Kotar, Nathan,
  Foffi, and Eiser}]{dimichele2013}
Di~Michele, L., F.~Varrato, J.~Kotar, S.~H. Nathan, G.~Foffi, and E.~Eiser,
  \enquote{Multistep kinetic self-assembly of dna-coated colloids,} Nature
  Communications \textbf{4}, 2007 (2013).

\bibitem[{Dibble \emph{et~al.}(2008)Dibble, Kogan, and Solomon}]{solomon}
Dibble, C.~J., M.~Kogan, and M.~J. Solomon, \enquote{Structural origins of
  dynamical heterogeneity in colloidal gels,} Phys. Rev. E \textbf{77}, 050401
  (2008).

\bibitem[{Dinsmore \emph{et~al.}(2006)Dinsmore, Prasad, Wong, and
  Weitz}]{dinsmore_prl2006}
Dinsmore, A.~D., V.~Prasad, I.~Y. Wong, and D.~A. Weitz, \enquote{Microscopic
  structure and elasticity of weakly aggregated colloidal gels,} Phys. Rev.
  Lett. \textbf{96}, 185502 (2006).

\bibitem[{Divoux \emph{et~al.}(2010)Divoux, Tamarii, Barentin, and
  Manneville}]{divoux2010transient}
Divoux, T., D.~Tamarii, C.~Barentin, and S.~Manneville, \enquote{Transient
  shear banding in a simple yield stress fluid,} Phys. Rev. Lett. \textbf{104},
  208301 (2010).

\bibitem[{Divoux \emph{et~al.}(2012)Divoux, Tamarii, Barentin, Teitel, and
  Manneville}]{divoux2012yielding}
Divoux, T., D.~Tamarii, C.~Barentin, S.~Teitel, and S.~Manneville,
  \enquote{Yielding dynamics of a herschel--bulkley fluid: a critical-like
  fluidization behaviour,} Soft Matter \textbf{8}(15), 4151--4164 (2012).

\bibitem[{Eberle \emph{et~al.}(2011)Eberle, Wagner, and Casta\~neda
  Priego}]{Eberle-prl2011}
Eberle, A. P.~R., N.~J. Wagner, and R.~Casta\~neda Priego, \enquote{Dynamical
  arrest transition in nanoparticle dispersions with short-range interactions,}
  Phys. Rev. Lett. \textbf{106}, 105704 (2011).

\bibitem[{Ewoldt \emph{et~al.}(2008)Ewoldt, Hosoi, and
  McKinley}]{Ewoldt-jor2008}
Ewoldt, R.~H., A.~E. Hosoi, and G.~H. McKinley, \enquote{New measures for
  characterizing nonlinear viscoelasticity in large amplitude oscillatory
  shear,} Journal of Rheology (1978-present) \textbf{52}(6) (2008).

\bibitem[{Falk and Langer(2011)}]{falk-langer}
Falk, M.~L., and J.~Langer, \enquote{Deformation and failure of amorphous,
  solidlike materials,} Annual Review of Condensed Matter Physics
  \textbf{2}(1), 353--373 (2011).

\bibitem[{Fall \emph{et~al.}(2010)Fall, Paredes, and Bonn}]{fall-prl2010}
Fall, A., J.~Paredes, and D.~Bonn, \enquote{Yielding and shear banding in soft
  glassy materials,} Phys. Rev. Lett. \textbf{105}, 225502 (2010).

\bibitem[{{Fielding}(2013)}]{fielding-arxiv2013}
{Fielding}, S.~M., \enquote{{Shear banding in soft glassy materials},} ArXiv
  e-prints, arXiv:1309.3422 [cond-mat.soft] (2013).

\bibitem[{Fielding and Olmsted(2003)}]{fielding-prl2003}
Fielding, S.~M., and P.~D. Olmsted, \enquote{Early stage kinetics in a unified
  model of shear-induced demixing and mechanical shear banding instabilities,}
  Phys. Rev. Lett. \textbf{90}, 224501 (2003).

\bibitem[{Fiocco \emph{et~al.}(2013)Fiocco, Foffi, and Sastry}]{fiocco-pre2013}
Fiocco, D., G.~Foffi, and S.~Sastry, \enquote{Oscillatory athermal quasistatic
  deformation of a model glass,} Phys. Rev. E \textbf{88}, 020301 (2013).

\bibitem[{Foffi \emph{et~al.}(2005)Foffi, De~Michele, Sciortino, and
  Tartaglia}]{foffi2005_jcp}
Foffi, G., C.~De~Michele, F.~Sciortino, and P.~Tartaglia, \enquote{Arrested
  phase separation in a short-ranged attractive colloidal system: A numerical
  study,} The Journal of chemical physics \textbf{122}, 224903 (2005).

\bibitem[{Gardel \emph{et~al.}(2004)Gardel, Shin, MacKintosh, Mahadevan,
  Matsudaira, and Weitz}]{gardel-science2004}
Gardel, M.~L., J.~H. Shin, F.~C. MacKintosh, L.~Mahadevan, P.~Matsudaira, and
  D.~A. Weitz, \enquote{Elastic behavior of cross-linked and bundled actin
  networks,} Science \textbf{304}(5675), 1301--1305 (2004).

\bibitem[{Gibaud \emph{et~al.}(2010)Gibaud, Frelat, and
  Manneville}]{gibaud2010heterogeneous}
Gibaud, T., D.~Frelat, and S.~Manneville, \enquote{Heterogeneous yielding
  dynamics in a colloidal gel,} Soft Matter \textbf{6}(15), 3482--3488 (2010).

\bibitem[{Gibaud and Schurtenberger(2009)}]{Gibaud-jpcm2009}
Gibaud, T., and P.~Schurtenberger, \enquote{A closer look at arrested spinodal
  decomposition in protein solutions,} Journal of Physics: Condensed Matter
  \textbf{21}(32), 322201 (2009).

\bibitem[{Gibaud \emph{et~al.}(2013)Gibaud, Zaccone, Del~Gado, Trappe, and
  Schurtenberger}]{gibaud-prl2013}
Gibaud, T., A.~Zaccone, E.~Del~Gado, V.~Trappe, and P.~Schurtenberger,
  \enquote{Unexpected decoupling of stretching and bending modes in protein
  gels,} Phys. Rev. Lett. \textbf{110}, 058303 (2013).

\bibitem[{Gisler \emph{et~al.}(1999)Gisler, Ball, and Weitz}]{gisler1999}
Gisler, T., R.~C. Ball, and D.~A. Weitz, \enquote{Strain hardening of fractal
  colloidal gels,} Phys. Rev. Lett. \textbf{82}, 1064--1067 (1999).

\bibitem[{Head \emph{et~al.}(2003)Head, Levine, and MacKintosh}]{head-pre2003}
Head, D.~A., A.~J. Levine, and F.~C. MacKintosh, \enquote{Distinct regimes of
  elastic response and deformation modes of cross-linked cytoskeletal and
  semiflexible polymer networks,} Phys. Rev. E \textbf{68}, 061907 (2003).


\bibitem[{Helgeson \emph{et~al.}(2014)Helgeson, Gao, Moran, Lee, Godfrin,
  Tripathi, Bose, and Doyle}]{Helgeson-softmat2014}
Helgeson, M.~E., Y.~Gao, S.~E. Moran, J.~Lee, M.~Godfrin, A.~Tripathi, A.~Bose,
  and P.~S. Doyle, \enquote{Homogeneous percolation versus arrested phase
  separation in attractively-driven nanoemulsion colloidal gels,} Soft matter
  \textbf{10}(17), 3122--3133 (2014).

\bibitem[{Heussinger and Frey(2006)}]{heussinger-prl2006}
Heussinger, C., and E.~Frey, \enquote{Floppy modes and nonaffine deformations
  in random fiber networks,} Phys. Rev. Lett. \textbf{97}, 105501 (2006).

\bibitem[{Hyun \emph{et~al.}(2011)Hyun, Wilhelm, Klein, Cho, Nam, Ahn, Lee,
  Ewoldt, and McKinley}]{Hyun-pps2011}
Hyun, K., M.~Wilhelm, C.~O. Klein, K.~S. Cho, J.~G. Nam, K.~H. Ahn, S.~J. Lee,
  R.~H. Ewoldt, and G.~H. McKinley, \enquote{A review of nonlinear oscillatory
  shear tests: Analysis and application of large amplitude oscillatory shear
  (laos),} Progress in Polymer Science \textbf{36}(12), 1697 -- 1753 (2011).

\bibitem[{Jamney \emph{et~al.}(2007)Jamney, McCormick, Rammensee, Leight,
  Georges, and MacKintosh}]{jamney-natmat2007}
Jamney, P.~A., M.~E. McCormick, S.~Rammensee, J.~L. Leight, P.~C. Georges, and
  F.~C. MacKintosh, \enquote{Negative normal stress in semiflexible biopolymer
  gels,} Nature Materials \textbf{6}(1), 48--51 (2007).

\bibitem[{Karmakar \emph{et~al.}(2010)Karmakar, Lerner, Procaccia, and
  Zylberg}]{procaccia}
Karmakar, S., E.~Lerner, I.~Procaccia, and J.~Zylberg, \enquote{Statistical
  physics of elastoplastic steady states in amorphous solids: Finite
  temperatures and strain rates,} Phys. Rev. E \textbf{82}, 031301 (2010).

\bibitem[{Kern and Frenkel(2003)}]{kern-frenkel}
Kern, N., and D.~Frenkel, \enquote{Fluid¬cfluid coexistence in colloidal
  systems with short-ranged strongly directional attraction,} The Journal of
  Chemical Physics \textbf{118}(21) (2003).


\bibitem[{Koumakis and Petekidis(2011)}]{petekidis_softmatter2011}
Koumakis, N., and G.~Petekidis, \enquote{Two step yielding in attractive
  colloids: transition from gels to attractive glasses,} Soft Matter
  \textbf{7}(6), 2456 (2011).

\bibitem[{Laurati \emph{et~al.}(2011)Laurati, Egelhaaf, and
  Petekidis}]{laurati_jor2011}
Laurati, M., S.~Egelhaaf, and G.~Petekidis, \enquote{Nonlinear rheology of
  colloidal gels with intermediate volume fraction,} J. Rheol. \textbf{55}(3),
  673 (2011).

\bibitem[{Laurati \emph{et~al.}(2009)Laurati, Petekidis, Koumakis, Cardinaux,
  Schofield, Brader, Fuchs, and Egelhaaf}]{exp2}
Laurati, M., G.~Petekidis, N.~Koumakis, F.~Cardinaux, A.~B. Schofield, J.~M.
  Brader, M.~Fuchs, and S.~U. Egelhaaf, J. Chem. Phys. \textbf{130}, 134907
  (2009).

\bibitem[{Lees and Edwards(1972)}]{leesedwards1972}
Lees, A., and S.~Edwards, \enquote{The computer study of transport processes
  under extreme conditions,} Journal of Physics C: Solid State Physics
  \textbf{5}(15), 1921 (1972).

\bibitem[{Lieleg \emph{et~al.}(2007)Lieleg, Claessens, Heussinger, Frey, and
  Bausch}]{lieleg-prl2007}
Lieleg, O., M.~M. A.~E. Claessens, C.~Heussinger, E.~Frey, and A.~R. Bausch,
  \enquote{Mechanics of bundled semiflexible polymer networks,} Phys. Rev.
  Lett. \textbf{99}, 088102 (2007).

\bibitem[{Lieleg \emph{et~al.}(2011)Lieleg, Kayser, Brambilla, Cipelletti, and
  Bausch}]{lieleg_nmat2011}
Lieleg, O., J.~Kayser, G.~Brambilla, L.~Cipelletti, and A.~R. Bausch,
  \enquote{Slow dynamics and internal stress relaxation in bundled cytoskeletal
  networks,} Nature Materials \textbf{10}(3), 236 (2011).

\bibitem[{Lindstrom \emph{et~al.}(2012)Lindstrom, Kodger, Sprakel, and
  Weitz}]{lindstrom-sm2012}
Lindstrom, S.~B., T.~E. Kodger, J.~Sprakel, and D.~A. Weitz,
  \enquote{Structures{,} stresses{,} and fluctuations in the delayed failure of
  colloidal gels,} Soft Matter \textbf{8}, 3657--3664 (2012).

\bibitem[{Lodge and Heyes(1999)}]{lodge-jor99}
Lodge, J. F.~M., and D.~M. Heyes, \enquote{Rheology of transient colloidal gels
  by brownian dynamics computer simulation,} Journal of Rheology (1978-present)
  \textbf{43}(1), 219--244 (1999).

\bibitem[{Lu \emph{et~al.}(2008)Lu, Zaccarelli, Ciulla, Schofield, Sciortino,
  and Weitz}]{plu-nature}
Lu, P.~J., E.~Zaccarelli, F.~Ciulla, A.~B. Schofield, F.~Sciortino, and D.~A.
  Weitz, \enquote{Gelation of particles with short-range attraction,} Nature
  \textbf{453}, 499 (2008).

\bibitem[{Maccarrone \emph{et~al.}(2010)Maccarrone, Brambilla, Pravaz, Duri,
  Ciccotti, Fromental, Pashkovski, Lips, Sessoms, Trappe, and
  Cipelletti}]{maccarrone}
Maccarrone, S., G.~Brambilla, O.~Pravaz, A.~Duri, M.~Ciccotti, J.~M. Fromental,
  E.~Pashkovski, A.~Lips, D.~Sessoms, V.~Trappe, and L.~Cipelletti,
  \enquote{Ultra-long range correlations of the dynamics of jammed soft
  matter,} Soft Matter \textbf{6}(21), 5514--5522 (2010).

\bibitem[{Maloney and Lema{\^\i}tre(2006)}]{maloney2006amorphous}
Maloney, C.~E., and A.~Lema{\^\i}tre, \enquote{Amorphous systems in athermal,
  quasistatic shear,} Physical Review E \textbf{74}(1), 016118 (2006).

\bibitem[{Martens \emph{et~al.}(2011)Martens, Bocquet, and
  Barrat}]{martens-prl2011}
Martens, K., L.~Bocquet, and J.-L. Barrat, \enquote{Connecting diffusion and
  dynamical heterogeneities in actively deformed amorphous systems,} Phys. Rev.
  Lett. \textbf{106}, 156001 (2011).

\bibitem[{Martin and Hu(2012)}]{martin2012transient}
Martin, J.~D., and Y.~T. Hu, \enquote{Transient and steady-state shear banding
  in aging soft glassy materials,} Soft Matter \textbf{8}(26), 6940--6949
  (2012).

\bibitem[{Masschaele \emph{et~al.}(2009)Masschaele, Vermant, and
  Fransaer}]{vermant_jor2009}
Masschaele, K., J.~Vermant, and J.~Fransaer, \enquote{Direct visualization of
  yielding in model two-dimensional colloidal gels subjected to shear flow,} J.
  Rheol. \textbf{53}, 1437 (2009).

\bibitem[{Milner(1993)}]{milner-pre93}
Milner, S.~T., \enquote{Dynamical theory of concentration fluctuations in
  polymer solutions under shear,} Phys. Rev. E \textbf{48}, 3674--3691 (1993).

\bibitem[{Mohraz and Solomon(2005)}]{Mohraz-jor2005}
Mohraz, A., and M.~J. Solomon, \enquote{Orientation and rupture of fractal
  colloidal gels during start-up of steady shear flow,} Journal of Rheology
  (1978-present) \textbf{49}(3) (2005).

\bibitem[{M\o{}ller \emph{et~al.}(2008)M\o{}ller, Rodts, Michels, and
  Bonn}]{moller2008shear}
M\o{}ller, P. C.~F., S.~Rodts, M.~A.~J. Michels, and D.~Bonn, \enquote{Shear
  banding and yield stress in soft glassy materials,} Phys. Rev. E \textbf{77},
  041507 (2008).

\bibitem[{Ohtsuka \emph{et~al.}(2008)Ohtsuka, Royall, and Tanaka}]{royall}
Ohtsuka, T., C.~P. Royall, and H.~Tanaka, \enquote{Local structure and dynamics
  in colloidal fluids and gels,} Europhys. Lett. \textbf{84}, 46002 (2008).

\bibitem[{Onuki(1992)}]{onuki}
Onuki, A., \enquote{Scattering from deformed swollen gels with
  heterogeneities,} J. Phys. II France \textbf{2}, 45 (1992).

\bibitem[{Ovarlez \emph{et~al.}(2009)Ovarlez, Rodts, Chateau, and
  Coussot}]{ovarlez-rheoacta2009}
Ovarlez, G., S.~Rodts, X.~Chateau, and P.~Coussot, \enquote{Phenomenology and
  physical origin of shear localization and shear banding in complex fluids,}
  Rheologica Acta \textbf{48}(8), 831--844 (2009).

\bibitem[{Pantina and Furst(2005)}]{pantina2005elasticity}
Pantina, J., and E.~Furst, \enquote{Elasticity and critical bending moment of
  model colloidal aggregates,} Phys. Rev. Lett. \textbf{94}(13), 138301 (2005).

\bibitem[{Picard \emph{et~al.}(2005)Picard, Ajdari, Lequeux, and
  Bocquet}]{picard-pre2005}
Picard, G., A.~Ajdari, F.~m.~c. Lequeux, and L.~Bocquet, \enquote{Slow flows of
  yield stress fluids: Complex spatiotemporal behavior within a simple
  elastoplastic model,} Phys. Rev. E \textbf{71}, 010501 (2005).

\bibitem[{Plimpton(1995)}]{plimpton1995fast}
Plimpton, S., \enquote{Fast parallel algorithms for short--range molecular
  dynamics,} J. Comp. Phys. \textbf{117}, 1--19 (1995).

\bibitem[{Pouzot \emph{et~al.}(2006)Pouzot, Nicolai, Benyahia, and
  Durand}]{Pouzot-jcis2006}
Pouzot, M., T.~Nicolai, L.~Benyahia, and D.~Durand, \enquote{Strain hardening
  and fracture of heat-set fractal globular protein gels,} Journal of Colloid
  and Interface Science \textbf{293}(2), 376 -- 383 (2006).

\bibitem[{Prasad \emph{et~al.}(2003)Prasad, Trappe, Dinsmore, Segre,
  Cipelletti, and Weitz}]{luca_faraday2003}
Prasad, V., V.~Trappe, A.~D. Dinsmore, P.~N. Segre, L.~Cipelletti, and D.~A.
  Weitz, \enquote{Rideal lecture. {U}niversal features of the fluid to solid
  transition for attractive colloidal particles,} Faraday Discuss.
  \textbf{123}, 1 (2003).

\bibitem[{Rajaram and Mohraz(2010)}]{rajaram_sm2010}
Rajaram, B., and A.~Mohraz, \enquote{Microstructural response of dilute
  colloidal gels to nonlinear shear deformation,} Soft Matter \textbf{6},
  2246--2259 (2010).

\bibitem[{Rovigatti \emph{et~al.}(2011)Rovigatti, Kob, and
  Sciortino}]{rovigatti2011}
Rovigatti, L., W.~Kob, and F.~Sciortino, \enquote{The vibrational density of
  states of a disordered gel model,} The Journal of chemical physics
  \textbf{135}, 104502 (2011).

\bibitem[{Sacanna \emph{et~al.}(2013)Sacanna, Korpics, Roriguez,
  Col\'on-Melendez, Kim, Pine, and Yi}]{sacanna2013}
Sacanna, S., M.~Korpics, K.~Roriguez, L.~Col\'on-Melendez, S.-Y. Kim, D.~J.
  Pine, and G.-R. Yi, \enquote{Shaping colloids for self-assembly,} Nature
  Communications \textbf{4}, 1688 (2013).

\bibitem[{Santos \emph{et~al.}(2013)Santos, Campanella, and
  Carignano}]{santos-sm2013}
Santos, P. H.~S., O.~H. Campanella, and M.~A. Carignano, \enquote{Effective
  attractive range and viscoelasticity of colloidal gels,} Soft Matter
  \textbf{9}, 709--714 (2013).

\bibitem[{Saw \emph{et~al.}(2009)Saw, Ellegaard, Kob, and
  Sastry}]{saw2009structural}
Saw, S., N.~Ellegaard, W.~Kob, and S.~Sastry, \enquote{Structural relaxation of
  a gel modeled by three body interactions,} Phys. Rev. Lett. \textbf{103}(24),
  248305 (2009).

\bibitem[{Schall and van Hecke(2009)}]{schall2009shear}
Schall, P., and M.~van Hecke, \enquote{Shear bands in matter with granularity,}
  Annual Review of Fluid Mechanics \textbf{42}(1), 67 (2009).

\bibitem[{Schall \emph{et~al.}(2007)Schall, Weitz, and
  Spaepen}]{schall-science2007}
Schall, P., D.~A. Weitz, and F.~Spaepen, \enquote{Structural rearrangements
  that govern flow in colloidal glasses,} Science \textbf{318}(5858),
  1895--1899 (2007).

\bibitem[{Schmitt \emph{et~al.}(1995)Schmitt, Marques, and
  Lequeux}]{schmitt-pre95}
Schmitt, V., C.~M. Marques, and F.~m.~c. Lequeux, \enquote{Shear-induced phase
  separation of complex fluids: The role of flow-concentration coupling,} Phys.
  Rev. E \textbf{52}, 4009--4015 (1995).

\bibitem[{Sciortino and Zaccarelli(2011)}]{sciortino}
Sciortino, F., and E.~Zaccarelli, \enquote{Reversible gels of patchy
  particles,} Current Opinion in Solid State and Materials Science
  \textbf{15}(6), 246--253 (2011).

\bibitem[{Seto \emph{et~al.}(2013)Seto, Botet, Meireles, Auernhammer, and
  Cabane}]{seto-jor2013}
Seto, R., R.~Botet, M.~Meireles, G.~K. Auernhammer, and B.~Cabane,
  \enquote{Compressive consolidation of strongly aggregated particle gels,}
  Journal of Rheology (1978-present) \textbf{57}(5), 1347--1366 (2013).

\bibitem[{Sottos and Moore(2011)}]{sottos_nature}
Sottos, N., and J.~S. Moore, \enquote{Material chemistry: Spot on healing,}
  Nature \textbf{472}, 299 (2011).

\bibitem[{Stillinger and Weber(1984)}]{stillinger-weber}
Stillinger, F.~H., and T.~A. Weber, \enquote{Packing structures and transitions
  in liquids and solids,} Science \textbf{225}(4666), 983--989 (1984).

\bibitem[{Storm \emph{et~al.}(2005)Storm, Pastore, MacKintosh, Lubensky, and
  Jamney}]{storm-nature2005}
Storm, C., J.~J. Pastore, F.~C. MacKintosh, T.~Lubensky, and P.~A. Jamney,
  \enquote{Nonlinear elasticity in biological gels,} Nature \textbf{435}(7039),
  191--194 (2005).

\bibitem[{Swan \emph{et~al.}(2014)Swan, Zia, and Brady}]{swan-jor2013}
Swan, J.~W., R.~N. Zia, and J.~F. Brady, \enquote{Large amplitude oscillatory
  microrheology,} Journal of Rheology (1978-present) \textbf{58}(1), 1--41
  (2014).

\bibitem[{Tanaka and Araki(2007)}]{tanaka2007_epl}
Tanaka, H., and T.~Araki, \enquote{Spontaneous coarsening of a colloidal
  network driven by self-generated mechanical stress,} EPL (Europhysics
  Letters) \textbf{79}(5), 58003 (2007).

\bibitem[{Tanguy \emph{et~al.}(2002)Tanguy, Wittmer, Leonforte, and
  Barrat}]{tanguy2002continuum}
Tanguy, A., J.~Wittmer, F.~Leonforte, and J.-L. Barrat, \enquote{Continuum
  limit of amorphous elastic bodies: A finite-size study of low-frequency
  harmonic vibrations,} Physical Review B \textbf{66}(17), 174205 (2002).

\bibitem[{Thompson \emph{et~al.}(2009)Thompson, Plimpton, and
  Mattson}]{thompson2009general}
Thompson, A.~P., S.~J. Plimpton, and W.~Mattson, \enquote{General formulation
  of pressure and stress tensor for arbitrary many-body interaction potentials
  under periodic boundary conditions,} The Journal of Chemical Physics
  \textbf{131}, 154107 (2009).

\bibitem[{Trappe \emph{et~al.}(2001)Trappe, Prasad, Cipelletti, Segre, and
  Weitz}]{trappe_nature2001}
Trappe, V., V.~Prasad, L.~Cipelletti, P.~Segre, and D.~Weitz, \enquote{Jamming
  phase diagram for attractive particles,} Nature \textbf{411}(6839), 772
  (2001).

\bibitem[{van~der Vaart \emph{et~al.}(2013)van~der Vaart, Rahmani, Zargar, Hu,
  Bonn, and Schall}]{vanderVaart-jor2013}
van~der Vaart, K., Y.~Rahmani, R.~Zargar, Z.~Hu, D.~Bonn, and P.~Schall,
  \enquote{Rheology of concentrated soft and hard-sphere suspensions,} Journal
  of Rheology (1978-present) \textbf{57}(4) (2013).

\bibitem[{Wyart \emph{et~al.}(2008)Wyart, Liang, Kabla, and
  Mahadevan}]{wyart-prl2008}
Wyart, M., H.~Liang, A.~Kabla, and L.~Mahadevan, \enquote{Elasticity of floppy
  and stiff random networks,} Phys. Rev. Lett. \textbf{101}, 215501 (2008).

\bibitem[{Yan \emph{et~al.}(2010)Yan, Altunbas, Yucel, Nagarkar, Schneider, and
  Pochan}]{pochan_softmatter2010}
Yan, C., A.~Altunbas, T.~Yucel, R.~P. Nagarkar, J.~P. Schneider, and D.~J.
  Pochan, \enquote{Injectable solid hydrogel: mechanism of shear-thinning and
  immediate recovery of injectable $\beta$-hairpin peptide hydrogels,} Soft
  Matter \textbf{6}, 5143 (2010).

\bibitem[{Zaccone \emph{et~al.}(2009)Zaccone, Wu, and
  Del~Gado}]{zaccone_prl2009}
Zaccone, A., H.~Wu, and E.~Del~Gado, \enquote{Elasticity of arrested
  short-ranged attractive colloids: Homogeneous and heterogeneous glasses,}
  Phys. Rev. Lett. \textbf{103}, 208301 (2009).

\end{thebibliography}

\end{document}